\documentclass[letterpaper]{article} 
\usepackage{aaai24}  
\usepackage{times}  
\usepackage{helvet}  
\usepackage{courier}  
\usepackage[hyphens]{url}  
\usepackage{graphicx} 
\usepackage{natbib}  
\usepackage{caption} 
\usepackage{algorithm}
\usepackage{algorithmic}
\usepackage{amsmath}
\usepackage{xcolor}
\usepackage{amsfonts}
\usepackage{booktabs}
\usepackage{multirow}
\usepackage{enumitem}
\usepackage[labelformat=simple]{subcaption}

\usepackage{xcolor}
\newcommand{\answerYes}[1]{\textcolor{blue}{#1}} 
\newcommand{\answerNo}[1]{\textcolor{teal}{#1}} 
\newcommand{\answerNA}[1]{\textcolor{gray}{#1}} 
 
\usepackage{newfloat}
\usepackage{listings}

\usepackage{newfloat}
\usepackage{listings}
\DeclareCaptionStyle{ruled}{labelfont=normalfont,labelsep=colon,strut=off} 
\floatstyle{ruled}
\newfloat{listing}{tb}{lst}{}
\floatname{listing}{Listing}
\title{Linguistic Landscape of Generative AI Perception: \\
A Global Twitter Analysis Across 14 Languages}

\author {
    Taichi Murayama\textsuperscript{\rm 1},
    Kunihiro Miyazaki\textsuperscript{\rm 2},
    Yasuko Matsubara\textsuperscript{\rm 3}, 
    Yasushi Sakurai\textsuperscript{\rm 3}
}
\affiliations {
    \textsuperscript{\rm 1}Yokohama National University\\
    \textsuperscript{\rm 2}Institute of Industrial Science, The University of Tokyo\\
    \textsuperscript{\rm 3}SANKEN, Osaka University\\
    murayama-taichi-bs@ynu.ac.jp, kunihirom@acm.com, \{yasuko, yasushi\}@sanken.osaka-u.ac.jp
}

\usepackage{bibentry}

\begin{document}

\maketitle

\begin{abstract}
The advent of generative AI tools has had a profound impact on societies globally, transcending geographical boundaries. 
Understanding these tools' global reception and utilization is crucial for service providers and policymakers in shaping future policies.
Therefore, to unravel the perceptions and engagements of individuals within diverse linguistic communities with regard to generative AI tools, we extensively analyzed over 6.8 million tweets in 14 different languages.
Our findings reveal a global trend in the perception of generative AI, accompanied by language-specific nuances.
While sentiments toward these tools vary significantly across languages, there is a prevalent positive inclination toward Image tools and a negative one toward Chat tools. 
Notably, the ban of ChatGPT in Italy led to a sentiment decline and initiated discussions across languages.
Furthermore, we established a taxonomy for interactions with chatbots, creating a framework for social analysis underscoring variations in generative AI usage among linguistic communities. 
We find that the Chinese community predominantly employs chatbots as substitutes for search, while the Italian community tends to use chatbots for tasks such as problem-solving assistance and engaging in entertainment or creative tasks.
Our research provides a robust foundation for further explorations of the social dynamics surrounding generative AI tools and offers invaluable insights for decision-makers in policy, technology, and education.
\end{abstract}

\section{Introduction}

Generative Artificial Intelligence (AI) has recently experienced a surge in popularity, representing significant advancements in various domains, including text generation, image creation, and software code development~\cite{brynjolfsson2023generative}. 
This growth has had an impact beyond the realm of technology, influencing various aspects of daily life~\cite{sallam2023chatgpt, lund2023chatgpt}, such as healthcare~\cite{gabashvili2023chatgpt} and education~\cite{bordt2023chatgpt, li2023chatgpt}. 
Notably, ChatGPT, with its user-friendly interface, has democratized access to AI, extending its usage from academics and professionals to the general public, and garnered one million users within just five days of its launch~\cite{five_days}.

The introduction of new services not only signifies technological advancements but also brings societal expectations and concerns, reminiscent of historical precedents such as the Luddite movement against mechanized looms (\citeauthor{luddite}) and the resistance against automated driving~\cite{corning}.
Gartner, Inc. emphasized in their press release that generative AI occupies the ``Peak of Inflated Expectations'' in their 2023 Hype Cycle for Emerging Technologies and is expected to yield transformative impacts in the coming years~\cite{gartner_press}.
In contrast, concerns about job displacement~\cite{PeopleFe45:online} and regulatory considerations, including bans in public schools~\cite{ChatGPTb11:online} and debates on ChatGPT's accessibility~\cite{rudolph6war,Germanyc2:online}, exist. 
To effectively address the challenges posed by generative AI in the future, it is imperative to comprehend societal perceptions and formulate strategies and policy guidelines for its widespread use.
Notably, generative AI primarily functions as an online service, transcending geographical boundaries and impacting nations worldwide. 
This global nature dictates diverse reactions and regulatory stances; for instance, some countries, such as Italy, have prohibited access to ChatGPT~\cite{italy_ban}, while others, such as Japan, are embracing its potential~\cite{openai_japan}.
These considerations underscore the importance of understanding how diverse cultures and languages perceive and utilize generative AI.

In this study, we conduct a comprehensive analysis of public perceptions and usage patterns of generative AI on Twitter (now X), especially focusing on the variations that emerge across distinct linguistic communities.
Given Twitter's extensive global user base and the frequent display of interactions with these technologies, it serves as an ideal platform for assessing public sentiments and identifying differences among languages. 
Understanding these distinctions is pivotal for considering the future dynamics between humans and AI. 
To this end, we pose these research questions:

\noindent \textbf{RQ1: How do the sentiments toward generative AI vary across different languages?} 
Our objective is to comprehend the trends and changes in the reception of generative AI within diverse linguistic communities. 
By examining the evolution of sentiments over time, we can grasp the nuanced differences in perception and ascertain how external factors relate to perceptions within various language communities.

\noindent \textbf{RQ2: How do linguistic communities differ in the content about generative AI tools?}
To gain a deeper insight into the content of tweets related to generative AI, we employ log-odds ratio and topic modeling.
These methods enable us to uncover varied perspectives and focal points that are indicative of distinct cultural and linguistic influences within these tweet contents.

\noindent \textbf{RQ3: How do people interact with chatbots?}
With a specific focus on ChatGPT, a prominent chatbot, we analyze the snapshots of interactions shared online. 
We employ topic modeling and open coding to reveal the nuanced disparities in user behaviors and intentions that are unique to each linguistic community when engaging with ChatGPT.

By answering these RQs, we made the following findings:
\begin{itemize}
    \setlength{\parskip}{0cm} 
    \setlength{\itemsep}{0cm} 
    \item Globally, there is a positive trend toward Image tools and a negative trend toward Chat tools. 
    These perceptions in the language community varied; \texttt{en} and \texttt{fr} communities, representing users tweeting in English and French respectively, displayed a higher sentiment as opposed to the lower sentiment exhibited by \texttt{ja}, \texttt{zh}, and \texttt{ru}. (Section \ref{sec4})

    \item A ban on ChatGPT in Italy and its prevalent usage have emerged as common topics worldwide in tweets. 
    Additionally, in each language community, we frequently observed discussions concerning NFTs and cryptocurrencies unique to the community. (Section \ref{sec5})    

    \item Through open coding, we developed a novel taxonomy, which clarifies the varied usage patterns of chatbots.
    For example, while the \texttt{zh} community predominantly employs chatbots as search alternatives, the \texttt{it} community tends to pose more complex prompts. (Section \ref{sec6})
\end{itemize}

This is the first comprehensive study to explore both user perspectives and user interactions with regard to generative AI across diverse linguistic communities.
Our findings unravel the linguistic nuances in the acceptance and utilization of generative AI.
Additionally, these insights can guide developers, policymakers, and educators in tailoring their strategies and initiatives to better align with specific cultural and linguistic communities.

\section{Related Work}
The public perception of AI has been the subject of extensive investigation, explored from various perspectives~\cite{kieslich2021threats, kelley2021exciting}.
The Pew Research Center has observed that many Americans harbor concerns concerning the integration of AI into their daily lives, particularly regarding issues related to employment and privacy~\cite{pew2022}.
Nevertheless, differential perception depending on the application of AI was noted~\cite{pew2023}.
For instance, AI's use in chemical elucidation is generally perceived as progressive, while its application in news writing is not seen as significantly progressive.
Zhang et al. conducted comprehensive interviews with Americans, revealing that the majority favored AI progression rather than against it~\cite{zhang2020us}.
Interview surveys to gauge people's perceptions of AI have been conducted across a diverse range of individuals and attributes, not limited to the U.S. 
These surveys span regions such as UK~\cite{cave2019scary}, India~\cite{kapania2022because}, and Australia~\cite{yigitcanlar2023artificial}, and communities such as US government employees~\cite{chen2023artificial}, teachers~\cite{nazaretsky2022instrument}, and researchers~\cite{chubb2022speeding}.
Apart from these survey-based studies, there have also been studies that attempt to understand perceptions toward AI from narrative stories~\cite{burgess2022rule, musa2020echoes} and news articles~\cite{nguyen2022news, karanouh2023mapping}.

Social media, with its capacity to provide real-time and extensive data, has proven to be a valuable tool for tracking dynamic trends and gauging public sentiments, as evidenced by its application in studying societal responses to technologies like IoT~\cite{bian2016mining}, self-driving cars~\cite{kohl2018anticipating} and AI~\cite{manikonda2018tweeting}.
Similarly, discussions surrounding generative AI have been abundant on these platforms~\cite{TheBrill64}, prompting researchers to utilize social media as a means of comprehending public perceptions regarding generative AI~\cite{haque2022think,qi2023excitements, rochadiani2023sentiment, korkmaz2023analyzing, taecharungroj2023can}.
For instance, Leiter et al. analyzed over 300,000 tweets and concluded that the hashtag ``\#ChatGPT'' predominantly garners positive feedback, characterized by joyful and favorable sentiments~~\cite{leiter2023chatgpt}.
Miyazaki et al. discovered that among various occupational groups, only illustrators showed negative sentiments toward generative AI, upon examining the relationship between perceptions and occupation groups~\cite{miyazaki2024public}.
Haensch et al.'s analysis of 100 ChatGPT-related TikTok videos revealed that a significant portion of these were tasks such as essay writing or coding~\cite{haensch2023seeing}.
However, a limitation of the current body of literature is its predominant focus on English-language posts. 
The diverse perceptions expressed across multiple languages and services regarding generative AI remain largely unexplored.

\section{Dataset}

\subsection{Target of Generative AI Tools}\label{3.1}
To generalize our research, we investigate various types of generative AI. 
Drawing from the enthusiasm observed on Twitter and insights derived from a prior study~\cite{miyazaki2024public}, we selected specific generative AI categorized based on their functionalities: conversational (Chat), image generation (Image), code completion (Code), and foundational models (Model).
These generative AI tools are listed in Table~\ref{target_ai}.
(Since they encompass both models and services of generative AI, we call them ``generative AI tools.'')

\begin{table}
  \small
  \caption{Targeted generative AI tools. Num (Orig.) indicates the initial tweet count, and Num (Fin.) indicates the remaining count after preprocessing and language identification, with the percentage of remaining tweets after data filtering shown in parentheses.
  }
  \vspace{-0.5em}
  \label{target_ai}
  \scalebox{0.81}{
  \begin{tabular}{cllrr}
    \toprule
    Category & Name & Release & Num (Orig.) & Num (Fin.)\\ \midrule
    \multirow{3}{*}{Chat} & ChatGPT & 11/30/2022& 5,981,521 & 4,376,767 (73.1\%)\\
    & Bing Chat & 02/22/2023 & 43,571  & 25,871 (59.4\%)\\
    & Perplexity AI & 12/07/2022& 5,959 & 5,249 (88.1\%)\\ \midrule
    \multirow{5}{*}{Image} & DALL·E & 07/20/2022& 500,248 & 410,862 (82.1\%)\\
    & Stable Diffusion & 01/05/2021& 867,695 & 572,160 (66.0\%)\\
    & Midjourney & 08/22/2022& 1,308,289 & 898,784 (68.7\%)\\ 
    & Craiyon & 04/21/2022& 167,590 & 84,797 (50.6\%)\\ 
    & DreamStudio & 08/20/2022 & 19,260 & 15,202 (78.9\%)\\ \midrule
    Code & Github Copilot & 10/29/2021 & 90,110 & 50,063 (55.6\%)\\ \midrule
    \multirow{3}{*}{Model} & GPT-3 & 06/11/2020 & 498,373 & 371,047 (74.5\%)\\
    & GPT-3.5 & 11/30/2022 & 39,140 & 25,204 (64.4\%)\\
    & GPT-4 & 03/14/2023 & 682,752 & 452,793 (66.3\%)\\ 
  \bottomrule
\end{tabular}
}
\vspace{-1em}
\end{table}

\subsection{Collecting of generative AI tweet}\label{3.2}
\noindent \textbf{Tweet retrieval.} 
We utilized the Twitter Academic API to extract tweets related to generative AI tools from June 11, 2020 (GPT-3's release), to May 2, 2023. 
We designed a comprehensive set of search keywords (details in Appendix B).
In total, we retrieved a total of 9,216,383 tweets, exclusive of retweets.
\looseness=-1

\noindent \textbf{Noise removal.}
To bolster the reliability of our subsequent analysis, we employed a three-step noise removal process on the collected tweets.
In particular, we removed tweets that meet these criteria: 1) tweets where the author's name or mentioned usernames include the name of the generative AI tool; 2) tweets discussing a generative AI tool before its release date; and 3) tweets originating from bot-like accounts (further elaborated in Appendix C).

\noindent \textbf{Language identification.}
Our primary objective is to discern disparities in perceptions and usage patterns of generative AI among diverse language communities.
To this end, we categorized the noise-filtered tweets according to their respective languages.
While most tweets inherently possess language tags, for those lacking explicit language tags, we employed lingua-py to determine the language of the text.
Our analysis is primarily focused on the 14 languages that appear most frequently within our collected tweets, removing tweets not written in these target languages.
The 14 languages are as follows: English (\texttt{en}), Japanese (\texttt{ja}), Spanish (\texttt{es}), French (\texttt{fr}), Portuguese (\texttt{pt}), Chinese (\texttt{zh}), German (\texttt{de}), Turkish (\texttt{tr}), Indonesian (\texttt{id}), Arabic (\texttt{ar}), Italian (\texttt{it}), Russian (\texttt{ru}), Korean (\texttt{ko}), and Dutch (\texttt{nl}).
In total, we obtained 6,864,568 tweets and identified 2,312,271 unique users in our dataset.

\subsection{Tweet data observations}\label{3.3}
\begin{table}
  \small
  \caption{
  Tweet volume and interest intensity by language for tweets related to generative AI tools. 
  The ``Number of Tweets'' column represents the total number of tweets about generative AI tools in each language, with the percentage of remaining tweets after data filtering shown in parentheses.
  }
  \label{language_statistics}
  \vspace{-0.5em}
  \begin{tabular}{rrrr}
    \toprule
    Language & Number of tweets & Number of users & IntI\\ \midrule
    English (\texttt{en}) & 4,053,410 (72.7\%) & 1,339,272 & 1.00\\ 
    Japanese (\texttt{ja}) & 1,463,957 (78.9\%) & 408,310 & 0.46\\ 
    Spanish (\texttt{es}) & 390,154 (77.1\%)& 164,720 & 0.41\\ 
    French (\texttt{fr}) & 249,855 (75.4\%)& 111,624 & 1.08\\ 
    Portuguese (\texttt{pt}) & 195,044 (83.6\%)& 113,650 & 0.23\\ 
    Chinese (\texttt{zh}) & 134,715 (66.4\%)& 44,217 & 1.67\\ 
    German (\texttt{de}) & 94,101 (78.4\%)& 40,349 & 0.94\\ 
    Turkish (\texttt{tr}) & 57,611 (87.7\%)& 32,061 & 0.17\\ 
    Indonesian (\texttt{id}) & 51,079 (65.6\%)& 33,105 & 1.12\\ 
    Arabic (\texttt{ar}) & 47,474 (82.1\%)& 22,637 & 0.14\\ 
    Italian (\texttt{it}) & 41,622 (69.9\%)& 22,453 & 0.41\\ 
     Russian (\texttt{ru}) & 29,006 (80.3\%)& 12,969 & 0.47\\ 
    Korean (\texttt{ko}) & 28,490 (61.6\%)& 10,302 & 0.07\\ 
    Dutch (\texttt{nl}) & 28,050 (90.7\%)& 15,625 & 0.56\\ 
  \bottomrule
\end{tabular}
\vspace{-1em}
\end{table}

To gauge the level of enthusiasm for generative AI tools across various languages, we introduce the ``Interest Intensity (IntI)'' index.
The index is derived from the average daily tweet volume for each language, aggregated over a three-month period in 2022 using Twitter sample stream data in \citeauthor{stream_tweet}.
A higher IntI value indicates a more active discussion of generative AI tools.
The index is calculated as follows:
\begin{align}
    \text{Interest } &\text{Intensity on target lang (IntI)} = \notag \\
    &\frac{
        \left(\frac{\text{Number of tweets about generative AI tools in target lang}}{\text{daily average tweets in target lang}}\right)
    }{
        \left(\frac{\text{Number of tweets about generative AI tools in \texttt{en}}}{\text{daily average tweets in \texttt{en}}}\right)
    }
\end{align}
Specifically, it reflects whether the proportion of tweets about these tools relative to daily tweets is greater or lesser than that in \texttt{en}.

Table~\ref{language_statistics} provides the number of tweets and the IntI for each language.
Notably, the \texttt{en} exhibits a higher IntI compared to most languages, indicating more active discussions regarding generative AI tools within this community.
However, \texttt{fr}, \texttt{id}, and \texttt{zh} have IntI values surpassing that of \texttt{en}.
Specifically, discussions about generative AI tools in \texttt{zh} occur at a frequency approximately 1.67 times greater than those in \texttt{en}.
The high level of interest in generative AI tools within the \texttt{zh} community is particularly notable, reflecting a strong engagement with emerging technologies
Conversely, in languages such as \texttt{ja} and \texttt{es}, where there is a substantial Twitter user base, the level of discussion is roughly 0.4 times that of \texttt{en}.

\section{RQ1: How do the sentiments toward generative AI vary across different languages?}\label{sec4}
We examine the temporal evolution of sentiments toward generative AI tools among various linguistic communities, highlighting differences in trends and shifts within these communities.

\begin{figure*}[t]
    \centering
    \includegraphics[width=18cm]{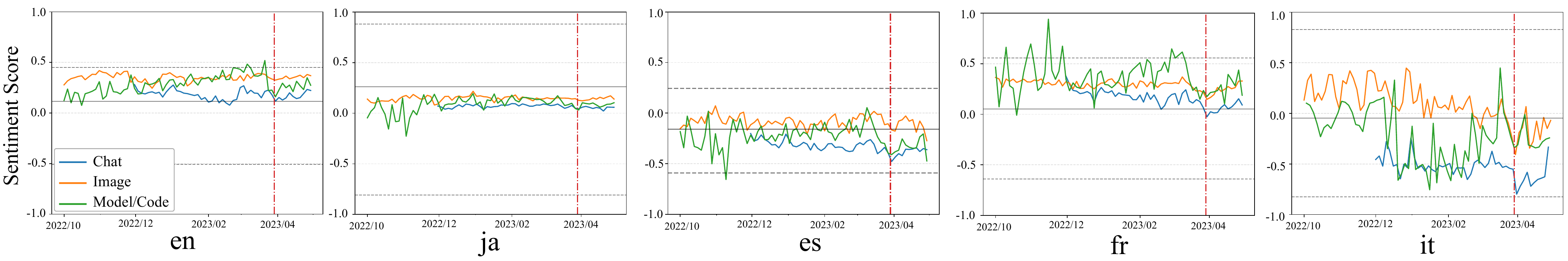}
    \caption{Time series of sentiment scores for five languages: \texttt{en}, \texttt{ja}, \texttt{es}, \texttt{fr}, and \texttt{it} from October 2022 to May 2, 2023.
    The x-axis represents time, and the y-axis represents the sentiment score. 
    The sentiment scores for generative AI tools are displayed as follows: the blue line represents Chat tool, orange represents Image tool, and green represents Model/Code tool.
    The solid gray line indicates the daily sentiment score, while the gray dotted lines represent the 75th and 25th percentiles of daily sentiment scores calculated from random tweets.
    The vertical red dotted line marks the date when Italy banned access to ChatGPT.
    We focus on the period from October 2022 onward due to the low volume of tweets and instability in sentiment data prior to this period, which made it difficult to visualize reliable trends across some languages.}
    \label{sentiment_trans}
\end{figure*}

\subsection{Sentiment classification model}
Despite the advancements in multilingual models~\cite{guo2020wiki, scao2022bloom, barbieri-etal-2022-xlm}, the task of conducting sentiment analysis across 14 languages using a unified model, as pursued in this section, remains challenging.
Consequently, we employed distinct monolingual sentiment classification models for the analysis of tweets in each language.
In our model selection process, we gave preference to publicly available models that have been trained on widely-used social media datasets such as SemEval-17~\cite{rosenthal-etal-2017-semeval}.
In cases wherein such models were unavailable, we selected prevalent models specific to the respective language. 
Additional details on the selected models for each language, along with their validation results, can be found in Appendix F.

These selected models evaluate each tweet, generating scores for both positive and negative sentiment classes.
To derive a single sentiment score for each tweet, the softmax function is applied to convert the two scores into probabilities and then calculate the ``sentiment score'' by subtracting the negative probability from the positive probability.
This score falls within the range of -1 to 1, with higher values indicating a more positive sentiment.

Given that the sentiment scores are generated using different models for each language, direct comparisons of sentiment scores across languages can introduce bias due to differences in model architectures and training datasets. 
To address this, our analysis focuses on how the sentiment scores about generative AI tools deviate from the daily sentiment in each language community, rather than comparing absolute sentiment scores across languages. 
To achieve this, we contrast the sentiment scores of tweets about generative AI tools with those of daily tweets in the same language. 
For this comparison, we randomly selected 50,000 tweets per language from the Twitter sample stream data in 2022 (\citeauthor{stream_tweet}) and calculated the daily average sentiment score. 
By comparing the sentiment scores for generative AI related tweets with these baseline daily sentiment scores, we aim to clarify whether perceptions of generative AI tools are more positive or negative compared to daily tweets within each language.

\begin{table}
  \small
  \caption{Sentiment score toward each category. The up and down arrows indicate whether the score is above or below the daily sentiment score for each language.}
  \label{sentiment_score_daily}
  \vspace{-0.5em}
  \begin{tabular}{crrrr}
    \toprule
    Language & Chat & Image & Model/Code & Daily\\ \midrule
    \texttt{en} & \textcolor{red}{$\uparrow \phantom{-}0.175$} & \textcolor{red}{$\uparrow \phantom{-}0.334$} & \textcolor{red}{$\uparrow \phantom{-}0.292$} & $0.116$\\ 
    \texttt{ja} & \textcolor{blue}{$\downarrow \phantom{-}0.064$} & \textcolor{blue}{$\downarrow \phantom{-}0.149$} & \textcolor{blue}{$\downarrow \phantom{-}0.101$} & 0.263\\ 
    \texttt{es} & \textcolor{blue}{$\downarrow -0.349$} & \textcolor{red}{$\uparrow -0.110$} & \textcolor{red}{$\uparrow -0.215$} & $-0.277$\\ 
    \texttt{fr} & \textcolor{red}{$\uparrow \phantom{-}0.130$} & \textcolor{red}{$\uparrow \phantom{-}0.280$} & \textcolor{red}{$\uparrow \phantom{-}0.326$} & $0.051$\\ 
    \texttt{pt} & \textcolor{blue}{$\downarrow -0.297$} & \textcolor{red}{$\uparrow -0.143$} & \textcolor{blue}{$\downarrow -0.247$} & $-0.174$\\ 
    \texttt{zh} & \textcolor{blue}{$\downarrow -0.095$} & \textcolor{blue}{$\downarrow -0.088$} & \textcolor{blue}{$\downarrow -0.131$} & $0.026$\\ 
    \texttt{de} & \textcolor{blue}{$\downarrow -0.118$} & \textcolor{red}{$\uparrow \phantom{-}0.070$} & \textcolor{blue}{$\downarrow -0.064$} & -0.029\\ 
    \texttt{tr} & \textcolor{blue}{$\downarrow \phantom{-}0.166$} & \textcolor{red}{$\uparrow \phantom{-}0.455$} & \textcolor{red}{$\uparrow \phantom{-}0.375$} & $0.222$\\ 
    \texttt{id} & \textcolor{blue}{$\downarrow -0.374$} & \textcolor{red}{$\uparrow \phantom{-}0.170$} & \textcolor{blue}{$\downarrow -0.250$} & $0.123$\\ 
    \texttt{ar} & \textcolor{blue}{$\downarrow \phantom{-}0.127$} & \textcolor{red}{$\uparrow \phantom{-}0.371$} & \textcolor{red}{$\uparrow \phantom{-}0.287$} & $0.286$\\ 
    \texttt{it} & \textcolor{blue}{$\downarrow -0.599$} & \textcolor{red}{$\uparrow \phantom{-}0.206$} & \textcolor{red}{$\uparrow \phantom{-}0.311$} & $-0.047$\\
    \texttt{ru} & \textcolor{blue}{$\downarrow -0.040$} & \textcolor{blue}{$\downarrow \phantom{-}0.140$} & \textcolor{blue}{$\downarrow -0.045$} & $0.379$\\ 
    \texttt{ko} & \textcolor{blue}{$\downarrow \phantom{-}0.199$} & \textcolor{red}{$\uparrow \phantom{-}0.352$} & \textcolor{red}{$\uparrow \phantom{-}0.305$} & $0.279$\\ 
    \texttt{nl} & \textcolor{blue}{$\downarrow \phantom{-}0.117$} & \textcolor{red}{$\uparrow \phantom{-}0.771$} & \textcolor{red}{$\uparrow \phantom{-}0.551$} & $0.531$\\ 
  \bottomrule
\end{tabular}
\vspace{-1.25em}
\end{table}

\subsection{Results and Findings}
Table~\ref{sentiment_score_daily} provides the sentiment scores by language and category.
A notable observation is that \texttt{en} and \texttt{fr} are the only languages in which the sentiment scores surpass the daily sentiment scores across all categories. 
This outcome may reflect greater familiarity with and acceptance of generative AI tools in these linguistic communities. 
In the case of \texttt{en}, this could be due to the fact that many generative AI tools originate in the U.S.
And, France is one of the more cautious European countries in regulating generative AI~\cite{french_europe}, indicating that the positive sentiments might indicate public support for the technology or effective communication about AI developments.
Conversely, \texttt{ja}, \texttt{zh}, and \texttt{ru} consistently score below the daily scores, signifying a more circumspect stance toward the emergence of generative AI tools in these linguistic communities. 
This may be due to cultural factors influencing public perception.

Another global trend is that, across the three categories, Chat tends to yield the lowest sentiment score, except in \texttt{zh} and \texttt{ru}, while Image consistently garners the highest sentiment score, except in \texttt{it} and \texttt{fr}. 
In the Chat category, only \texttt{en} and \texttt{fr} exhibit sentiment scores higher than the daily scores, which aligns with the increased discussion surrounding regulation after the global proliferation of chat-based generative AI. 
This indicates a contrasting sentiment between chat tools and image tools.

Several hypotheses may explain this contrast. 
First, differences in the user groups interested in each tool may play a role.
Despite image tools appearing earlier than chat tools, we observed fewer tweets about image tools compared to chat tools (see Section 3). 
This suggests that image tools were primarily discussed by users with a strong interest in technology or early adopters, who may have been more enthusiastic and positive due to their novelty and creative possibilities. 
In contrast, chat tools like ChatGPT gained widespread mainstream attention, leading to a rapid increase in related tweets from not only tech enthusiasts but also the general public.
This broader exposure included users who might feel anxious about job security and ethical implications, possibly revealing negative sentiments.
Secondly, differences in functionalities may contribute to the contrasting sentiments. Image tools are often used for creative expression and recreation, leading to enjoyment and personal satisfaction, which may foster positive sentiments. 
Conversely, chat tools are frequently used in professional settings for task management, customer service, and other work-related activities. 
The functionality of chat tools may cause people to consider the possibility of job displacement. 
Indeed, prominent words in the negative sentiment toward chat tools include ``destroy'' and ``dangerous,'' which are not commonly seen with other tools, suggesting fears of disruption to existing norms and concerns over potential negative impacts on society (refer to Appendix G.)

Figure~\ref{sentiment_trans} illustrates the time series of sentiment scores for five languages: \texttt{en}, \texttt{ja}, \texttt{es}, \texttt{fr}, and \texttt{it}.
Of these, \texttt{ja} consistently displays a stable sentiment. 
This stability can be attributed to Japan's relatively fewer regulations or policies, which allow for regular discussions on the features and various aspects of generative AI tools, with minimal influence from external events.
For \texttt{it}, the sentiment score for Chat tools has consistently hovered around -0.5 since their release. 
This particular release has notably influenced to the perception of other generative AI tools, resulting in a further decline in sentiment scores for Model/Code and Image. 
One significant event is the sentiment drop on March 29, affecting multiple languages, especially those in Europe, coinciding with Italy's ban on ChatGPT~\cite{italy_ban}.
Remarkably, the sentiment scores for Chat, Image, and Model/Code all experienced a simultaneous decline, underscoring the significant impact of this event. 
These insights underscore the importance of early sentiment assessment for emerging tools across different languages, providing critical perspectives for tool providers aiming to refine their promotional strategies.

\begin{figure*}[t]
    \centering
    \caption{Top 10 words more likely to be used by odds ratio, except generative AI tool' names and seasonal words, in six languages.
  Red indicates words that appear in multiple languages and blue indicates words unique to a specific language.}
  \vspace{-0.5em}
    \includegraphics[width=18.2cm]{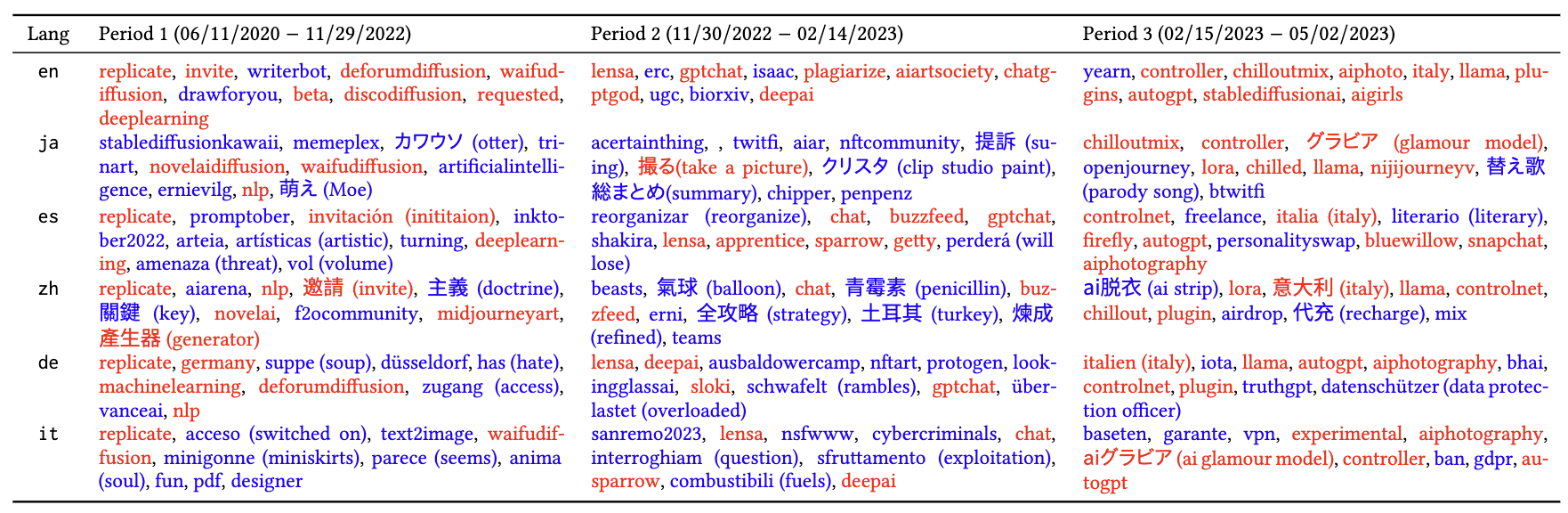}
    \label{odds_ratio_table}
    \vspace{-1.5em}
\end{figure*}

\section{RQ2: How do linguistic communities differ in the content about generative AI tools?}\label{sec5}
Gaining an understanding of how people reference generative AI tools, the expressions they use, and the topics they touch upon provides valuable insights into its global perception. 
To explore these aspects, we employ two distinct methodologies: odds ratio analysis and topic modeling.
Odds ratio analysis facilitates the quantitative measurement of the frequency of specific terms, shedding light on the common words employed when discussing this technology. 
Simultaneously, topic modeling facilitates the exploration of the underlying topics and prevalent themes within discussions about generative AI tools.

\subsection{Odds ratio analysis}
\noindent \textbf{Methodology: }
To reveal the dynamics of word significance across three distinct time intervals, we performed an odds ratio analysis employing logistic regression.
We segmented our dataset based on ChatGPT's release, a pivotal event that considerably influenced the popularity of generative AI tools. 
The data were categorized into three distinct periods: before the emergence of ChatGPT ($06/11/2020 - 11/29/2022$; Period 1), the early period after the emergence of ChatGPT ($11/30/2022 - 2/14/2023$; Period 2), and the late period after the emergence of ChatGPT ($02/15/2023 - 05/2/2023$; Period 3).
We then transformed each word into a count vector and employed a multinomial logistic regression model to calculate the odds ratio of each word by exponentiating the model's coefficients~\cite{sperandei2014understanding}.

\noindent \textbf{Results and Findings: } 
Figure~\ref{odds_ratio_table} presents the top 10 words for six language communities ranked by their odds ratios (see Figure 15 in the Appendix for a complete list).
In Period 1, we observed significant mentions of ``replicate,'' an experimentation platform for generative AI, and  ``invite,'' reflecting early restricted access to DALL·E and Midjourney.
There is evident active usage of pre-prompted notebooks such as ``deforumdiffusion,'' ``waifudiffusion,'' and ``novelaidiffusion,'' suggesting substantial user engagement in image generation.
Additionally, words associated with core generative AI technologies such as ``deeplearning,'' ``nlp,'' and ``machinelearning'' exhibit high odds ratios, signifying a pronounced focus on fundamental technologies during this period.
In Period 2, words related to ChatGPT such as ``chatgptgod'' and ``gptchat'' are frequently observed.
Interestingly, the word ``buzzfeed'' exhibits high odds ratios across multiple languages, owing to their plans with ChatGPT~\cite{buzzfeedmake}.
Furthermore, discussions on generative AI often coincide with debates on NFTs and cryptocurrencies, as indicated by terms such as ``erc,'' ``isaac,'' ``nftart,'' ``penpenz,'' and ``sloki.''
Period 3 features mentions of ``controller,'' ``plugins,'' and ``autogpt,'' indicating a creative engagement with ChatGPT prompts. 
The emergence of new services and models such as ``llama,'' ``lora,'' adobe ``firefly,'' and ``chillout'' is also notable. 
Remarkably, words related to ``italy'' are prevalent in all languages except \texttt{ja} and \texttt{ko}, underlining the significant impact of the ban on ChatGPT.
As demonstrated above, a common trend is discerned to some extent globally.
Words appearing across multiple languages accounted for 39.8\% of the top 10 words in Period 1, but 60.7\% of those in Period 3, suggesting that discussions about generative AI tools are gradually becoming language independent and global.

Language-specific patterns are also evident.
The \texttt{ja} community spotlight local AI services including ``memeplex,'' ``clip studio paint,'' and ``nijijourney,'' demonstrating domestic tech growth.
The \texttt{es} community in Period 1 exhibits a growing interest in an online image generation event called ``promptber,'' along with the appearance of the term ``artistic,'' indicating an active engagement in image generation.
Conversely, the appearance of terms such as ``threat'' presents an intriguing aspect. 
In the \texttt{it} community, discussions revolve around ``vpn'' and the General Data Protection Regulation (``gdpr’'), likely triggered by the ChatGPT ban, illustrating heightened awareness of regulatory impacts.
These observations underscore the nuanced interactions and unique reactions of different linguistic communities within the evolving landscape of generative AI.

\begin{table*}[t]
    \footnotesize
    \caption{Keywords for the top five topics and topic number in terms of number of tweets in each language by BERTopic. Red indicates topics that appear across multiple languages, and blue indicates topics specific to that language that do not appear in the top 50 topics of other languages.}
    \vspace{-1em}
\begin{subtable}{1.0\textwidth}
    \captionsetup{justification=centering, labelsep=period}
    \caption{Topics on tweet contents}
    \vspace{-0.7em}
    \scalebox{0.75}{
    \begin{tabular}{clclclcl} \toprule
    \multirow{5}{*}{\texttt{en}} & \textcolor{red}{0: chatgpt, chatgptcool, doolally} &
    \multirow{5}{*}{\texttt{ja}} & \textcolor{blue}{46: kun, chan, atsuto} &
    \multirow{5}{*}{\texttt{es}} & \textcolor{red}{1: bing, search, edge} & 
    \multirow{5}{*}{\texttt{fr}} & 28: midjourney, midjourneydiscord, lifesuck \\
    
    & 2: marketcap, zsp, zenithswap & & 
    \textcolor{red}{1: bing, search, edge} & &
    18: chatgptbugbbbcnews, what, about & & 
    \textcolor{blue}{113: french, cnil, macron} \\
    
    & \textcolor{red}{1: bing, search, edge} & & 
    \textcolor{blue}{31: excel, vba, spreadsheet} & &
    \textcolor{blue}{175: industrial, azure, offer} & & 
    19: midjourney, v5, hirasawa 
    \\
    
    & 9: exceeds, limit, character & &
    16: recipe, ingredients, cooking & & 
    \textcolor{red}{8: italy, privacy, ban} & & 
    \textcolor{red}{1: bing, search, edge}
    \\
    
    & 3: dalle, generation, waifulabs & &
    54: nijijourney, touhou, aiar & & 
    202: repeatedly, requests, waiting & & 
    \textcolor{red}{8: italy, privacy, ban}
    \\ 
    \midrule

     \multirow{5}{*}{\texttt{pt}} & 6: chatgpt, please, professional &
     \multirow{5}{*}{\texttt{zh}} & 50: chatgpt, summary, channel & 
     \multirow{5}{*}{\texttt{de}} & \textcolor{red}{11: gpt4, thank, user} & 
     \multirow{5}{*}{\texttt{tr}} & \textcolor{red}{11: gpt4, thank, user}
     \\

     & \textcolor{red}{1: bing, search, edge} & & 
     \textcolor{red}{1: bing, search, edge} & & 
     \textcolor{blue}{508: nft, read2n, mins} & & 
     \textcolor{red}{8: italy, privacy, ban}
     \\

     &  \textcolor{red}{8: italy, privacy, ban} & & 
     \textcolor{blue}{141: mainland, chinese, gao} & & 
     \textcolor{red}{1: bing, search, edge} & & 
     \textcolor{blue}{603: turkish, republic, kurdish}
     \\
     
     &  7: chatgpt, work, home & & 
     \textcolor{blue}{432: airdrop, usdc, pump} & & 
     \textcolor{red}{4: tweet, threads, timeline} & & 
     110: exam, pass, studying
     \\

     & 56: music, lyric, song & & 
     \textcolor{blue}{438: bounght, coin, fired} & & 
     \textcolor{red}{8: italy, privacy, ban} & & 
     \textcolor{red}{1: bing, search, edge}
     \\ 
     \midrule

     \multirow{5}{*}{\texttt{id}} & \textcolor{red}{11: gpt4, thank, user} &
     \multirow{5}{*}{\texttt{ar}} & \textcolor{blue}{673: activation, vpn, whatsapp} & 
     \multirow{5}{*}{\texttt{it}} & \textcolor{red}{11: gpt4, thank, user} & 
     \multirow{5}{*}{\texttt{ru}} & \textcolor{red}{0: chatgpt, chatgptcool, doolally}
     \\

     & 305: cryptovoxels, chatgpt, gpt3 & & 
     \textcolor{red}{11: gpt4, thank, user} & & 
     \textcolor{blue}{477: guarantor, privacy, italy} & & 
     \textcolor{blue}{1772: networkpainting, neural, telegram}
     \\

     & 14: gpt3, smiled, looked & & 
     \textcolor{blue}{894: activate, password, whatsapp} & & 
     20: photo, image, visualgpt & & 
     95: ukraine, russia, putin
     
     \\
     
     & \textcolor{blue}{340: galxeoat, brg, uler} & & 
     \textcolor{red}{8: italy, privacy, ban} & & 
     181: unfortunately, yeah, alright & & 
     \textcolor{red}{8: italy, privacy, ban}
     \\

     & 51: craiyon, disappoint, formerly & & 
     \textcolor{blue}{1228: arabic, calligraphy, harami} & & 
     86: benchmark, hallucinate, gpt35 & & 
     \textcolor{red}{4: tweet, threads, timeline}
     \\
     \midrule
     
     \multirow{5}{*}{\texttt{ko}} & \textcolor{blue}{692: laughing, humor, dalza} & 
     \multirow{5}{*}{\texttt{nl}} & \textcolor{red}{11: gpt4, thank, user} 
     \\

     & \textcolor{blue}{284: korea, takeshima, declare} & & 
     33: teacher, classroom, education
     \\
     
     & \textcolor{red}{1: bing, search, edge} &&
     \textcolor{red}{8: italy, privacy, ban}
     \\

     & 24: search, engine, google & & 
     \textcolor{red}{4: tweet, threads, timeline}
     \\

     & 60: translate, deeply, google & & 
     \textcolor{red}{1: bing, search, edge}
     \\
    \bottomrule
    \end{tabular}
    \vspace{-1.5em}
    }
    \label{bertopic_result_rq2}
\end{subtable}

\bigskip 

\begin{subtable}{1.0\textwidth}
    \vspace{-0.5em}
    \captionsetup{justification=centering, labelsep=period}
    \caption{Topics on prompts}
    \vspace{-0.7em}
    \scalebox{0.75}{
    \begin{tabular}{clclclcl} \toprule
    \multirow{5}{*}{\texttt{en}} & \textcolor{red}{0: answer, short, english} &
    \multirow{5}{*}{\texttt{ja}} & \textcolor{blue}{2: translate, japanese, sentence} &
    \multirow{5}{*}{\texttt{es}} & 9: load, span, mobile & 
    \multirow{5}{*}{\texttt{fr}} & \textcolor{red}{3: tweet, buzz, funny} \\
    
    & \textcolor{red}{3: tweet, buzz, funny} & & 
    5: tell, profile, who & &
    \textcolor{blue}{77: chile, peru, president} & & 
    \textcolor{red}{1: chatgpt, gpt4, chat} \\
    
    & \textcolor{red}{7: poet, rhyme, write} & & 
    \textcolor{red}{4: game, play, board} & &
    \textcolor{red}{0: answer, short, english} & & 
    \textcolor{red}{16: joke, humor, women} 
    \\
    
    & \textcolor{red}{1: chatgpt, gpt4, chat} & &
    \textcolor{red}{1: chatgpt, gpt4, chat} & & 
    \textcolor{red}{11: intelligence, ai, human} & & 
    \textcolor{blue}{125: france, pension, retirement}
    \\
    
    & \textcolor{red}{16: joke, humor, women} & &
    \textcolor{red}{6: medical, patient, treatment} & & 
    \textcolor{red}{1: chatgpt, gpt4, chat} & & 
    \textcolor{red}{11: intelligence, ai, human}
    \\ 
    \midrule

     \multirow{5}{*}{\texttt{pt}} & \textcolor{blue}{62: president, brazil, election} &
     \multirow{5}{*}{\texttt{zh}} & \textcolor{blue}{105: chinese, zhou, qing} & 
     \multirow{5}{*}{\texttt{de}} & \textcolor{red}{3: tweet, buzz, funny} & 
     \multirow{5}{*}{\texttt{tr}} & \textcolor{blue}{222: ataturk, turkey, ottoman}
     \\

     & \textcolor{red}{0: answer, short, english} & & 
     \textcolor{blue}{44: card, alipay, wechat} & & 
     19: students, teachers, education & & 
     \textcolor{blue}{376: turkeys, jealous, turkishness}
     \\

     &  \textcolor{blue}{181: football, gerais, brazil} & & 
     \textcolor{blue}{130: baidu, weibo, suspend} & & 
     \textcolor{red}{1: chatgpt, gpt4, chat} & & 
     \textcolor{blue}{603: turkish, republic, kurdish}
     \\
     
     &  \textcolor{red}{16: joke, humor, women} & & 
     \textcolor{blue}{221: xi, jinping, dictator} & & 
     \textcolor{blue}{253: austria, chancellor, german} & & 
     \textcolor{blue}{555: fenerbahe, football, champion}
     \\

     & \textcolor{blue}{230: portuguese, brazilian, accent} & & 
     45: price, nasdaq, stock & & 
     73: authoritarian, cannabis, prohibition & & 
     \textcolor{blue}{265: earthquake, disaster, seismic}
     \\ 
     \midrule

     \multirow{5}{*}{\texttt{id}} & \textcolor{blue}{283: indonesia, jakarta, malaysia} &
     \multirow{5}{*}{\texttt{ar}} & \textcolor{blue}{290: arabic, abdullah, bin} & 
     \multirow{5}{*}{\texttt{it}} & \textcolor{blue}{150: italy, disabled, guarantor} & 
     \multirow{5}{*}{\texttt{ru}} & 79: ukraina, russia, war
     \\

     & \textcolor{red}{1: chatgpt, gpt4, chat} & & 
     \textcolor{red}{6: medical, patient, treatment} & & 
     \textcolor{blue}{477: italy, privacy, blocked} & & 
    \textcolor{red}{16: joke, humor, women}
     \\

     & \textcolor{red}{6: medical, patient, treatment} & & 
     203: openai, available, country & & 
     \textcolor{red}{0: answer, short, english} & & 
     \textcolor{red}{3: tweet, buzz, funny}
     
     \\
     
     & 14: image, photo, camera & & 
     \textcolor{red}{11: intelligence, ai, human} & & 
     \textcolor{blue}{807: resume, connect, pleased} & & 
     \textcolor{red}{4: game, play, board}
     \\

     & 548: \textcolor{blue}{islam, shahada, muslim} & & 
     64: occasionally, quantum, limitations & & 
     \textcolor{blue}{429: denied, access, site} & & 
     \textcolor{blue}{1131: mongolian, kyrgyzstan, atadan}
     \\
     \midrule
     
     \multirow{5}{*}{\texttt{ko}} & \textcolor{blue}{167: korea, north, president} & 
     \multirow{5}{*}{\texttt{nl}} & 18: vaccine, virus, covid19
     \\

     & \textcolor{blue}{347: korea, heungbujeon, hangul} & & 
     \textcolor{red}{3: tweet, buzz, funny}
     \\
     
     & \textcolor{red}{4: game, play, board} &&
     \textcolor{red}{1: chatgpt, gpt4, chat}
     \\

     & \textcolor{blue}{417: kim, lee, hwan} & & 
     \textcolor{red}{7: poet, rhyme, write}
     \\

     & \textcolor{red}{6: medical, patient, treatment} & & 
     13: gender, lgbt, sexual
     \\
    \bottomrule
    \end{tabular}
    }
    \label{bertopic_result_rq3}
\end{subtable}
\end{table*}

\subsection{Topic modeling}\label{topic_model_rq2}

\noindent \textbf{Methodology: }
To uncover prevalent discussions both globally and unique to specific language communities about generative AI tools, we employed a topic model.
Applying a singular topic model to our dataset enables the identification of common discussions across different languages. 
For this purpose, we translated all tweets into English by using Google Translate. 
For tweets originally in English, a back-translation through Spanish is conducted to ensure consistency~\cite{klotz2023back}. 
Despite the ongoing debate about utilizing translation for multilingual content analysis~\cite{thompson2019ensuring, balahur2012multilingual, araujo2020comparative}, several studies affirm that combining topic models with machine translation effectively captures the main themes of text~\cite{maier2022machine,lind2022building}.

Topic modeling encompasses a diverse range of methods, including probabilistic models like Latent Dirichlet Allocation (LDA)~\cite{chauhan2021topic}, neural-based topic models~\cite{grootendorst2022bertopic}, graph-based approaches~\cite{yang2020graph}, and others.
Each of these methods has its strengths and applications depending on the nature of the dataset and the goals of the analysis.
In our study, we opted for BERTopic~\cite{grootendorst2022bertopic}, a neural topic modeling approach that leverages embeddings and clustering algorithms. 
BERTopic is well-suited for our dataset, which consists of short texts like tweets, as it excels at capturing semantic relationships and producing coherent topics from brief inputs~\cite{ebeling2022analysis}. 
Unlike traditional models like LDA, which treat words as independent entities, BERTopic embeds words in a high-dimensional space, allowing for more nuanced and contextually aware topic extraction.
Moreover, given that all tweets in our dataset were translated into English, using a model like BERTopic, which captures semantic similarities through embeddings, was critical. 
The translation process can lead to subtle changes in word choice or phrasing, but BERTopic mitigates these challenges by focusing on the underlying meaning rather than surface-level word frequencies. 
This made BERTopic the optimal choice for identifying key topics and patterns across language communities with greater accuracy.\footnote{One could also consider using a multilingual pre-training model instead of relying on translation.
When applying BERTopic with a multilingual pre-training model, it was observed that most topics with the highest frequency were specific to each language, and no common topics emerged across different languages. A comparison of the results of translating multilingual text into English and applying BERTopic (English BERTopic) with the results of applying multilingual BERTopic is shown in the Appendix H.}

\noindent \textbf{Results and Findings: }
Upon applying BERTopic to all translated tweets, 54.1\% are categorized as outliers, with the remaining tweets classified under 4,420 distinct topics.
Table~\ref{bertopic_result_rq2} presents the keywords for the top five topics per language.
Given the large number of topics, each topic comprises less than 1\% of the tweets in each language.

Several topics appear consistently across multiple languages.
Notably, discussions on the integration of search engines and generative AI (Topic 1), regulations related to ChatGPT in Italy (Topic 8), and expressions of admiration for ChatGPT and GPT-4 (Topics 0, 11) were prominent.
These topics emerged organically from our exploratory analysis, reflecting global events and user interests.

Our analysis also revealed unique discussions specific to particular language communities. 
For example, we observed connections between generative AI and local events (Topic 113 in \texttt{fr}, Topic 141 in \texttt{zh}, Topic 603 in \texttt{tr}, Topic 1228 in \texttt{ar}, Topic 603 in \texttt{zh}, Topic 95 in \texttt{ru}, and Topic 284 in \texttt{ko}.)
These localized discussions often reflect cultural or regional concerns about generative AI.
Moreover, topics related to NFTs and cryptocurrencies appeared prominently across several language communities. 
However, there were variations in the specific currencies and NFTs discussed, leading to unique language-specific topics (Topics 432 and 438 in \texttt{zh}, Topic 340 in \texttt{id}, Topic 508 in \texttt{de}.)
In the \texttt{ja} community, Topic 46, which includes terms such as ``kun’’ and ``chan’’ (equivalent to Mr. and Ms.), emerges from the practice of appending honorific titles to these tools, such as \texttt{ChatGPT-kun}, showing how these tools are integrated into everyday discourse in Japan.
The \texttt{zh} and \texttt{ar} communities in the Orient feature a majority of top topics specific to their communities, indicating distinct discussions compared to others.
Particularly noteworthy is the \texttt{ar} community's high-ranking topics (673 and 894), which highlight a trend in selling GPT accounts via social platforms, a phenomenon that is less prevalent in other language communities.

Our analysis reveals both universally relevant topics and those specific to particular languages or cultures, enhancing our understanding of generative AI discourse within diverse linguistic contexts.~\footnote{For a more detailed analysis of the relationships between topics within each language community, please refer to Appendix I, where we explore the topic networks and interconnections specific to each linguistic group.}
Furthermore, it is important to note that the prominence of ChatGPT-related topics is largely a reflection of the high volume of tweets about ChatGPT across various language communities.

\section{RQ3: How do people interact with chatbots?}\label{sec6}
The emergence of chatbots has garnered significant attention, leading to their increasing adoption and utilization in our daily lives. 
In addressing this RQ, we aimed to elucidtate how chatbots, with a primary focus on ChatGPT, are being utilized by a diverse population from various linguistic backgrounds.
To this end, we clarify the distinctions among language communities from two perspectives: by examining the topic of prompts using the topic model and by analyzing the utilization of chatbots using open coding.

\subsection{Extraction interactions from images}
Numerous users share screenshots of their ChatGPT interactions on Twitter, serving as a valuable resource for understanding its usage~\cite{TheBrill64}. 
Since most users do not include their prompts or responses in the tweet text, extracting text from screenshots is essential to capture the full scope of user interactions. This method is both cost-effective and scalable compared to alternatives like surveys or interviews.

We employ a two-step approach to analyze these resources as text data: 1) Rule-based classification of ChatGPT screenshots from images, and 2) Conversion of the extracted images into text, including prompts and responses, using Optical Character Recognition. 
Through these steps, we successfully retrieved a total of 507,714 interactions with ChatGPT, referred to as the ``Interaction dataset.'' 
For a consistent analysis, these interactions were translated into English.

\subsection{Topic modeling}
Topic modeling serves as an effective tool for understanding the prompts employed by each language community.
As discussed in Section \ref{sec5}, we applied BERTopic to the prompts of all interactions, resulting in 53.9\% of them being categorized as outliers, with the remaining interactions classified under 1,518 topics.
Table~\ref{bertopic_result_rq3} shows the keywords for the top five topics per language.
There are recurrent topics across multiple languages that revolve around prompts for search and generation; searching for chatbot usage (Topic 1), games (Topic 4), and medical information (Topic 6); generating tweets (Topic 3), poetry (Topic 7), and jokes (Topic 16).
Language-specific observations reveal prominent language-specific prompts related to political topics (Topic 77 in \texttt{es}, Topic 125 in \texttt{fr}, Topic 62 in \texttt{pt}, Topic 221 in \texttt{zh}, Topic 603 in \texttt{tr}, Topic 290 in \texttt{ar}, and Topic 167 in \texttt{ko}) and related to sports topics (Topic 181 in \texttt{pt} and Topic 555 in \texttt{tr}) in the major country of the language community.
Especially, the \texttt{zh} and \texttt{tr} communities use unique prompts related to social issues in these communities not found in other languages, accentuating the distinctiveness of these communities.
In the \texttt{it} community, as in the other results, the influence of the ChatGPT ban was evident in popular prompts.

\begin{table}
    \footnotesize
    \centering
    \caption{Taxonomy of main categories for chatbot usage.}
    \vspace{-0.7em}
    \scalebox{0.80}{
    \begin{tabular}{r p{3cm} p{4cm} r} 
    \toprule
        No. & Category Name & Description & Num (Pct.)\\ \midrule
        1 & \textbf{Search} & Utilizing chatbots to retrieve specific information akin to web search & 2,335 (35.6\%)\\ 
        2 & \textbf{Questions or Requests Reflecting the Responder's Preferences} & Engaging chatbots to obtain personalized or subjective responses & 1,727 (26.4\%)\\ 
        3 & \textbf{Support for Business or Creative Tasks} & Leveraging chatbots to provide support in business or creative tasks & 1,440 (22.0\%)\\
        4 & \textbf{Dialogue} & Engaging in conversational interactions with chatbots & 414 (6.3\%)\\
        5 & \textbf{Humor, Wit, Riddles} & Engaging with chatbots for amusement, humor, or intellectual challenge & 319 (4.9\%)\\
        6 & \textbf{Other} & Various other interactions or inquiries that do not fall under the above categories & 318 (4.9\%)\\
    \bottomrule
    \end{tabular}
    }
    \vspace{-1em}
    \label{taxonomy_chat_main}
\end{table}

\subsection{Classification based on open coding}
We sought to understand the content of interactions through topic modeling, however, the specific purposes for which chatbots are utilized remains an open issue.
This open issue motivates us to formulate a detailed taxonomy using the open coding technique~\cite{glaser1968discovery}.
Open coding, a fundamental step in qualitative research, entails the process of categorizing raw data into distinct codes (categories).
This method has previously been applied to derive categories from online data, such as sexism~\cite{samory2021call}, hate speech~\cite{salminen2018anatomy}, and anxiety about COVID-19~\cite{chen2023you}.
We aim to establish a taxonomy for chatbot usage by employing open coding, annotate a subset of the data, and then employ machine learning to automate the classification.

\subsubsection{Building codebook}
Guided by open coding principles~\cite{glaser1968discovery, corbin1990grounded}, we initiated the categorization process using randomly selected samples from the Interaction dataset. 
The first author categorized the samples until reaching a saturation point, where no new categories were emerging. 
This saturation point was achieved after processing 300 samples, a number deemed reasonable based on previous studies~\cite{maxwell2020short, chen2023you}.
Throughout this process, we refined and expanded categories, and merged certain subcategories into larger ones to enhance clarity and coherence.
Ultimately, the first author organized chatbot usage into main categories and subcategories. 
Each sample was consistently assigned to a main category and, where applicable, a subcategory, which was devised to group samples frequently found within the main category.
The devised categories and associated samples underwent review by the other authors, and through discussions, we built a codebook.

\subsubsection{Taxonomy and Annotation}
We identified six main categories and 13 subcategories of chatbot usage.
Then, to annotate these categories based on the constructed taxonomy, we randomly selected 500 samples for each language.
After filtering out noise, a total of 6,553 samples were independently annotated by five annotators. 
The inter-annotator agreement was assessed in 500 samples using Cohen's Kappa coefficient, resulting in values of 0.842 for the main categories and 0.712 for all categories, including the subcategories.
This high level of agreement instills confidence in the annotated dataset, which we refer to as the ``Chatbot usage dataset.''
Descriptions of the main categories and their label distributions are shown in Table~\ref{taxonomy_chat_main}.~\footnote{Observations and results for the subcategories are presented in the Appendices K, M, and O. However, because of limited accuracy in all category classifications, the analyses and discussions in this paper exclusively focus on the main categories.}


\subsubsection{Classification Model}\label{simple_classification}
To achieve automatic categorization of abundant interactions, we constructed classification models of main categories from interactions in the Chatbot usage dataset.
We developed four different models and evaluated their accuracy: TF-IDF + Logistic Regression (LR), GPT-3 (DaVinci variant~\cite{brown2020language},) BERT~\cite{kenton2019bert} (pre-trained model of \cite{bert_uncased},) and DeBERTa~\cite{he2020deberta} (pre-trained model of \cite{microsoft_debert}.)
The dataset was partitioned into training, validation, and testing sets in a 7:1:2 ratio, respectively. 

\begin{table}
    \small
    \centering
    \caption{Main category classification performance; This reports P: precision, R: recall, F1-score of positive class on the utilization for each category and  Macro and Weighted Avg for all samples.}
    \vspace{-0.7em}
    \scalebox{0.62}{
    \begin{tabular}{rllllllllllll} \toprule
          & \multicolumn{3}{c}{TF-IDF + LR} & \multicolumn{3}{c}{GPT-3} & \multicolumn{3}{c}{BERT} & \multicolumn{3}{c}{DeBERTa}\\
         Variable & P & R & F1 & P & R & F1 & P & R & F1 & P & R & F1 \\
         \midrule
         1 Search & 0.65 & 0.53 & 0.58 & 0.74 & 0.67 & 0.70 & 0.71 & 0.82 & 0.76 & 0.78 & 0.74 & 0.76\\
         2 Preference & 0.47 & 0.50 & 0.49 & 0.60 & 0.65 & 0.62 & 0.65 & 0.54 & 0.60 & 0.69 & 0.73 & 0.71 \\
         3 Support & 0.60 & 0.71 & 0.65 & 0.65 & 0.65 & 0.62 & 0.81 & 0.69 & 0.74 & 0.79 & 0.74 & 0.76 \\
         4 Dialogue & 0.59 & 0.65 & 0.62 & 0.65 & 0.84 & 0.73 & 0.42 & 0.46 & 0.44 & 0.69 & 0.62 & 0.65 \\
         5 Humor & 0.61 & 0.72 & 0.66 & 0.45 & 0.35 & 0.39 & 0.38 & 0.45 & 0.41 & 0.62 & 0.75 & 0.68\\
         6 Other & 0.42 & 0.51 & 0.39 & 0.57 & 0.45 & 0.51 & 0.51 & 0.57 & 0.54 & 0.38 & 0.49 & 0.42\\ \midrule
         Macro Avg & 0.56 & 0.60 & 0.57 & 0.60 & 0.54 & 0.56 & 0.58 & 0.59 & 0.58 & 0.66 & 0.68 & 0.66\\
         Weighted Avg & 0.59 & 0.58 & 0.58 & 0.65 & 0.65 & 0.64 & 0.67 & 0.66 & 0.66 & 0.72 & 0.73 & 0.72 \\
         \bottomrule
    \end{tabular}
    }
    \vspace{-0.5em}
    \label{main_category}
\end{table}

Table~\ref{main_category} presents the results of the classification in the main category of chatbot usage.
The performances of the GPT-3 and BERT models exhibit high F1 scores exceeding 0.60 in categories with a substantial number of samples (1: Search, 2: Preference, and 3: Support); however, they show low F1 scores in categories with relatively fewer samples.
The TF-IDF + LR model consistently achieves F1 scores around 0.60, regardless of the sample size within categories, although its overall accuracy falls short compared to the other models.
The DeBERTa model emerges as the most balanced, offering commendable F1 scores even in small sample sizes (4: Dialogue and 5: Humor.)
Consequently, we selected the DeBERTa model for subsequent analysis.

\begin{figure}
    \centering
    \includegraphics[width=9cm]{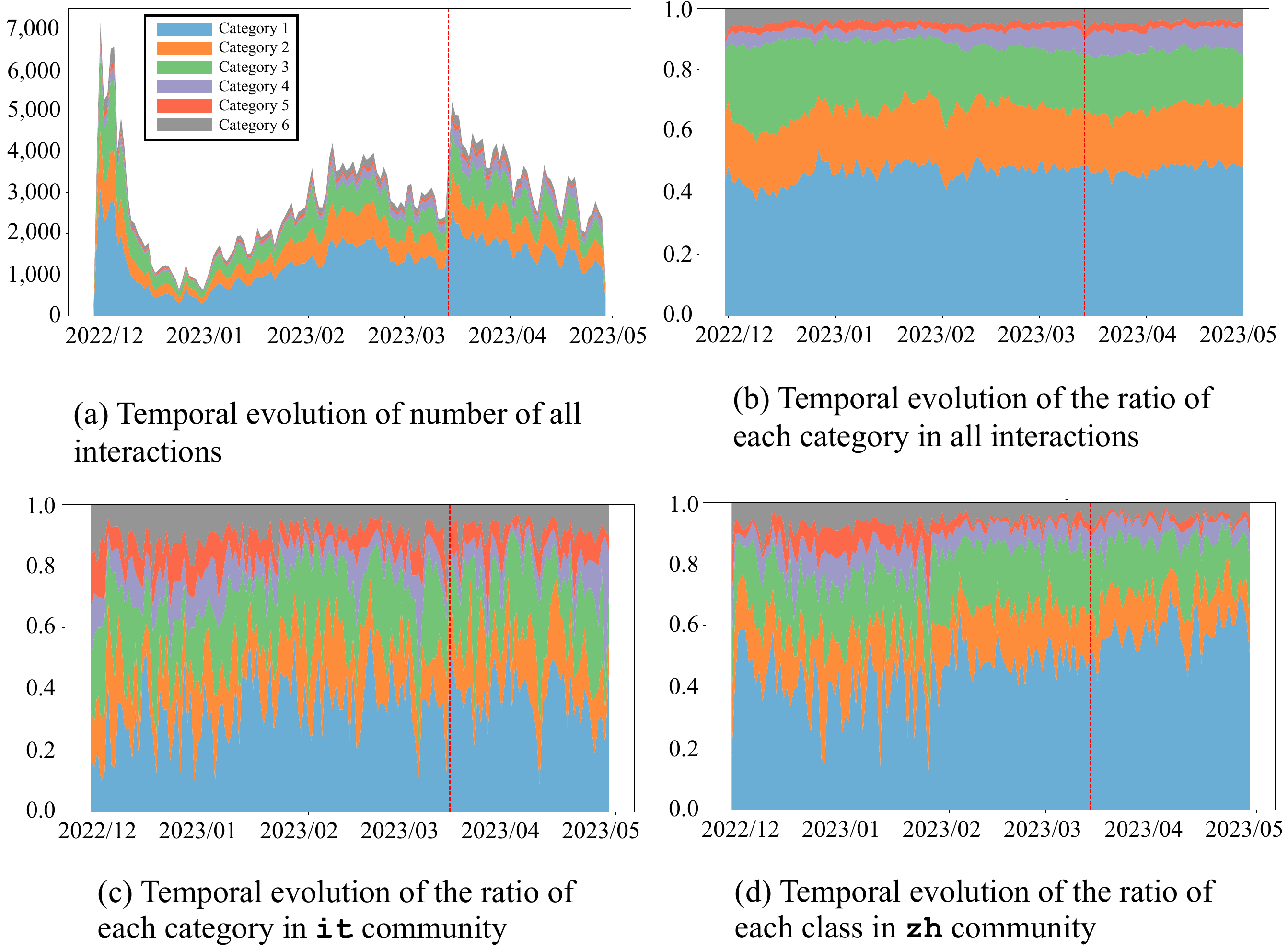}
    \vspace{-0.9em}
    \caption{Temporal evolution of chatbot usage across different categories. The vertical red line indicates the release date of GPT-4. The x-axis represents time, and the y-axis represents number or ratio of interactions in each category.}
    \label{timeseries_classification}
    \vspace{-1em}
\end{figure}

\subsubsection{Analysis on the evolution of chatbot usage}
We apply the aforementioned DeBERTa model to all samples in the Interaction dataset for the classification of the main category, aiming to understand the transition in chatbot usage over time.
In Figure~\ref{timeseries_classification} (a), which shows the number of all interactions, a significant peak was observed just after the release of ChatGPT, followed by a gradual slowdown, but the number of interactions started to grow again after January 2023.
The initial surge in tweets can be attributed to the early adopters who first took notice of the service. 
Although there was a subsequent decline, the trend reversed as the service gained global traction, leading to an increase in tweets related to chatbot usage.
Figure~\ref{timeseries_classification} (b), showing the ratio of each category, indicates that the ratio does not deviate significantly from the ratio of the annotated data.
A comparison between December 2022 (immediately after ChatGPT's release) and April 2023 shows an increase in Category 4 (Dialogue), which points toward a growing engagement in more conversational interactions over time.

Looking at different language communities, we found unique usage patterns for chatbots. 
For example, Figure~\ref{timeseries_classification} (c) shows that the \texttt{it} community has more interactions in Categories 3 (Support) and 5 (Humor) compared to other language communities. 
This pattern suggests that the \texttt{it} community uses ChatGPT for assistance with tasks or problem-solving (3 Support) and for entertainment or playful exchanges (5 Humor), not just for straightforward information retrieval tasks (1 Search.)
When juxtaposed with RQ1 findings, there is a noticeable relationship between higher engagement in ``Support'' and ``Humor'' interactions and less favorable sentiment toward chatbots, possibly stemming from not meeting expectations in these tasks. 
The \texttt{zh} community, by contrast, evidenced a steady increase in Category 1 (Search) as per Figure~\ref{timeseries_classification} (d).
This community shows a stronger growing trend in using ChatGPT for straightforward information retrieval, indicating its widespread use as a search tool.

\section{Limitations and Future work}
\noindent \textbf{Discrepancy between social media perceptions and actual usage:}
Social media serves as a valuable medium for assessing societal reactions and perceptions toward generative AI. 
While the data reflect certain aspects of these usages, they may not provide a complete picture of real-world usage, especially in the case of chatbot usages as investigated in RQ3. 
In other words, the interactions shown on social media may represent only a fraction of actual usage scenarios.
However, social media enables large-scale investigations that might be infeasible through traditional survey methods. 
Our work, leveraging data from over 500,000 interactions, provides a foundational understanding and a categorization of chatbot usage, laying the groundwork for future explorations. 
Despite its limitations in capturing the full extent of reality, social media remains a valuable data source for preliminary research.

\noindent \textbf{Keyword exhaustiveness:}
Using English-centric tool names or Wikipedia as a source for keyword selection, while practical, may overlook the given names of generative AI tools in a particular language.
The challenge of keyword exhaustiveness underscores the importance of a well-thought-out strategy for keyword selection to achieve a comprehensive analysis.

\noindent \textbf{Coverage and bias of generative AI tools:}
Our research, conducted in May 2023, focuses on generative AI tools that had been discussed on social media up to that point. 
In our analysis, likely, some generative AI tools were not covered, and the notable excitement surrounding ChatGPT may have also resulted in a biased analysis outcome toward chatbots.
Although our study aimed to understand the differences in perceptions across language communities concerning generative AI tools, a more granular analysis, grouping posts by specific services and analyzing topics, could offer a more nuanced understanding of the public perceptions and usages of generative AI.
Additionally, as newer generative AI models such as LLaMA, Bard, and Gemini have emerged and gained popularity after May, 2023, it will be essential for future studies to include these tools. This will ensure that analyses remain up-to-date and reflective of the evolving landscape of AI technologies.

\noindent \textbf{Representation limitations of linguistic communities:}
One of the inherent challenges in analyzing social media data across multiple languages is the varying levels of representation among linguistic communities. 
In some cases, the number of tweets analyzed represents only a small subset of the broader language community, and the insights gathered may not fully represent the entire user base of those language communities. 
This raises the question of whether the findings truly reflect the opinions and behaviors of the entire linguistic community. 
Additionally, even within dominant languages like English and Japanese, certain demographics, such as specific age groups or regional users, may be overrepresented or underrepresented, introducing potential biases.
These limitations are inherent in many social media studies, where the user base may not fully reflect the broader population. 
While our study aims to provide insights into the perceptions of generative AI tools on Twitter, highlighting both language-specific phenomena and globally common trends, it is important to recognize these representational limitations for interpreting the broader applicability and scope of our findings.

\section{Conclusions}
Through our exploration of user perceptions and interactions with generative AI tools across diverse linguistic communities, we uncovered both shared global patterns and unique usages and themes specific to each community.
These findings not only contribute to academic discussions on human-AI interaction but also offer a comprehensive understanding of how these innovative tools are received and utilized within various cultural and linguistic contexts.
Moreover, our analysis sheds light on the diverse dynamics of social engagement with generative AI tools, particularly in how different communities employ them for distinct purposes. 
For instance, Chinese users primarily use chatbots as search engines, whereas Italian users tend to engage with chatbots for creative writing or problem-solving. 
These differences highlight the significant role that cultural and social contexts play in shaping user expectations and interactions with AI.
Our insights provide valuable guidance for shaping strategies and initiatives that are tailored to the specific cultural and linguistic nuances of each community, ultimately promoting a more inclusive and effective deployment of generative AI technologies.

\section*{Ethical Considerations}
The data in this paper is derived from publicly-accessible user-generated content.
We pay the utmost attention to the privacy of individuals in this study. We did not discuss the results regarding specific users to keep the privacy of the individual studied in this paper.
\looseness=-1

\section*{Acknowledgement}
This work was partly supported by
JST CREST JPMJCR23M3,
JSPS KAKENHI Grant-in-Aid for Scientific Research Number 23K16889, and 
Research Institute of Science and Technology for Society, Japan, Grant Number JPMJRS23L4.

\bibliography{aaai24}

\section*{Paper Checklist}

\begin{enumerate}

\item For most authors...
\begin{enumerate}
    \item  Would answering this research question advance science without violating social contracts, such as violating privacy norms, perpetuating unfair profiling, exacerbating the socio-economic divide, or implying disrespect to societies or cultures?
    \answerYes{Yes}
  \item Do your main claims in the abstract and introduction accurately reflect the paper's contributions and scope?
    \answerYes{Yes}
   \item Do you clarify how the proposed methodological approach is appropriate for the claims made? 
     \answerYes{Yes}
   \item Do you clarify what are possible artifacts in the data used, given population-specific distributions?
     \answerYes{Yes}
  \item Did you describe the limitations of your work?
    \answerYes{Yes}
  \item Did you discuss any potential negative societal impacts of your work?
    \answerYes{Yes}
      \item Did you discuss any potential misuse of your work?
    \answerYes{Yes}
    \item Did you describe steps taken to prevent or mitigate potential negative outcomes of the research, such as data and model documentation, data anonymization, responsible release, access control, and the reproducibility of findings?
    \answerYes{Yes, see Ethical Considerations}
  \item Have you read the ethics review guidelines and ensured that your paper conforms to them?
    \answerYes{Yes}
\end{enumerate}

\item Additionally, if your study involves hypotheses testing...
\begin{enumerate}
  \item Did you clearly state the assumptions underlying all theoretical results?
    \answerNA{NA}
  \item Have you provided justifications for all theoretical results?
    \answerNA{NA}
  \item Did you discuss competing hypotheses or theories that might challenge or complement your theoretical results?
    \answerNA{NA}
  \item Have you considered alternative mechanisms or explanations that might account for the same outcomes observed in your study?
    \answerNA{NA}
  \item Did you address potential biases or limitations in your theoretical framework?
   \answerNA{NA}
  \item Have you related your theoretical results to the existing literature in social science?
    \answerNA{NA}
  \item Did you discuss the implications of your theoretical results for policy, practice, or further research in the social science domain?
    \answerNA{NA}
\end{enumerate}

\item Additionally, if you are including theoretical proofs...
\begin{enumerate}
  \item Did you state the full set of assumptions of all theoretical results?
    \answerNA{NA}
	\item Did you include complete proofs of all theoretical results?
    \answerNA{NA}
\end{enumerate}

\item Additionally, if you ran machine learning experiments...
\begin{enumerate}
  \item Did you include the code, data, and instructions needed to reproduce the main experimental results (either in the supplemental material or as a URL)?
    \answerYes{Yes, see Appendices}
  \item Did you specify all the training details (e.g., data splits, hyperparameters, how they were chosen)?
    \answerYes{Yes, see Appendix N}
     \item Did you report error bars (e.g., with respect to the random seed after running experiments multiple times)?
    \answerNA{NA}
	\item Did you include the total amount of compute and the type of resources used (e.g., type of GPUs, internal cluster, or cloud provider)?
    \answerNo{No}
     \item Do you justify how the proposed evaluation is sufficient and appropriate to the claims made? 
    \answerYes{Yes}
     \item Do you discuss what is ``the cost`` of misclassification and fault (in)tolerance?
    \answerYes{Yes, see Section 6}
  
\end{enumerate}

\item Additionally, if you are using existing assets (e.g., code, data, models) or curating/releasing new assets, \textbf{without compromising anonymity}...
\begin{enumerate}
  \item If your work uses existing assets, did you cite the creators?
    \answerYes{Yes}
  \item Did you mention the license of the assets?
    \answerYes{Yes}
  \item Did you include any new assets in the supplemental material or as a URL?
    \answerYes{Yes}
  \item Did you discuss whether and how consent was obtained from people whose data you're using/curating?
    \answerYes{Yes, see Ethical Considerations}
  \item Did you discuss whether the data you are using/curating contains personally identifiable information or offensive content?
    \answerYes{Yes, see Ethical Considerations}
\item If you are curating or releasing new datasets, did you discuss how you intend to make your datasets FAIR (see \citet{fair})?
\answerYes{Yes}
\item If you are curating or releasing new datasets, did you create a Datasheet for the Dataset (see \citet{gebru2021datasheets})? 
\answerNA{NA}
\end{enumerate}

\item Additionally, if you used crowdsourcing or conducted research with human subjects, \textbf{without compromising anonymity}...
\begin{enumerate}
  \item Did you include the full text of instructions given to participants and screenshots?
    \answerYes{see, Appendix L}
  \item Did you describe any potential participant risks, with mentions of Institutional Review Board (IRB) approvals?
    \answerNA{NA}
  \item Did you include the estimated hourly wage paid to participants and the total amount spent on participant compensation?
    \answerNA{NA}
   \item Did you discuss how data is stored, shared, and deidentified?
   \answerYes{Yes}
\end{enumerate}
\end{enumerate}

\newpage
\appendix

\section{Global Policy Landscape on Generative AI}
In the rapidly evolving landscape of generative AI, nations worldwide are adapting and implementing policies to ensure the technology's responsible development and deployment~\cite{hutson2023rules, promise, hacker2023regulating, cath2018artificial, djeffal2022role, fatima2020national}.

In the U.S., Executive Order 13960 from 2021, titled ``Promoting the Use of Trustworthy AI in the Federal Government,'' was issued to enhance public trust in the AI applications of federal agencies~\cite{eo13960}.
In October 2022, further emphasizing its commitment to AI development, the White House Office of Science and Technology Policy (OSTP) released the ``Blueprint for an AI Bill of Rights,’’ outlining five key principles for AI utilization to address various challenges associated with the technology~\cite{ai_bill}.
In 2023, with the growing recognition of the risks associated with generative AI, the National Institute of Standards and Technology (NIST), an agency under the U.S. Department of Commerce, released the Artificial Intelligence Risk Management Framework 1.0 (AI RMF) in January. 
This framework serves as a voluntary, non-sector-specific, use-case-agnostic guide for technology companies~\cite{tabassi2023artificial}.
Highlighting the concerns surrounding AI's inherent risks, the White House, in collaboration with OSTP, formulated a strategy in May 2023 to ensure fairness and transparency in AI technology deployment~\cite{biden_administration, response_ostp}.
With government agencies regulating as well as influential figures such as Elon Musk calling for a halt to AI development due to the risks to society of AI~\cite{elon_musk_ai}, the U.S. has become the center of the generative AI debate.

In the UK, the government published a white paper expressing its supportive stance toward AI innovation ~\cite{uk_government_approach}.
This document underscores the importance of adapting existing frameworks to incorporate AI considerations while avoiding overly restrictive regulations that could hinder innovation. 
Conversely, Italy took a different approach by temporarily banning the use of ChatGPT at the end of March, citing the need for measures to protect personal data~\cite{italy_ban}. 
In parallel, the adoption of an ``AI Act’’ was adopted on March 13, 2024 in the European Union (EU) region, against a background of the proliferation of generative AI~\cite{eu_ac_act}. 
This act categorizes AI applications based on their risk levels, aiming to mitigate potential hazards and prohibit high-risk uses, including real-time biometric surveillance in public areas.

China has proactively taken steps to regulate generative AI, introducing regulations in November 2022 to ensure the security of providers of deep synthesis technology~\cite{china_deep}.
Additionally, the government has enforced a ban on access to chatbot-based services such as ChatGPT because of concerns about the safety of content generated by chatbots~\cite{china_ban}.
In August 2023, the Cyberspace Administration of China (CAC) issued administrative measures aimed at monitoring and controlling generated content and safeguarding personal information~\cite{china_aug}.
In contrast, Japan lacks clear regulations or policies for generative AI and relies primarily on guidelines and warnings~\cite{kojinjyoho}.

Policies regarding generative AI differ across nations and cultural contexts~\cite{fatima2021explains}.
When formulating AI regulations, it is crucial to tailor policies to the distinct political and economic landscapes of individual cultures rather than seeking a one-size-fits-all global standard~\cite{nitzberg2022algorithms, gianni2022governance}.

\begin{figure}[t]
    \centering
    \caption{List of search keywords of generative AI tools}
    \label{search_keyword}
    \includegraphics[width=9cm]{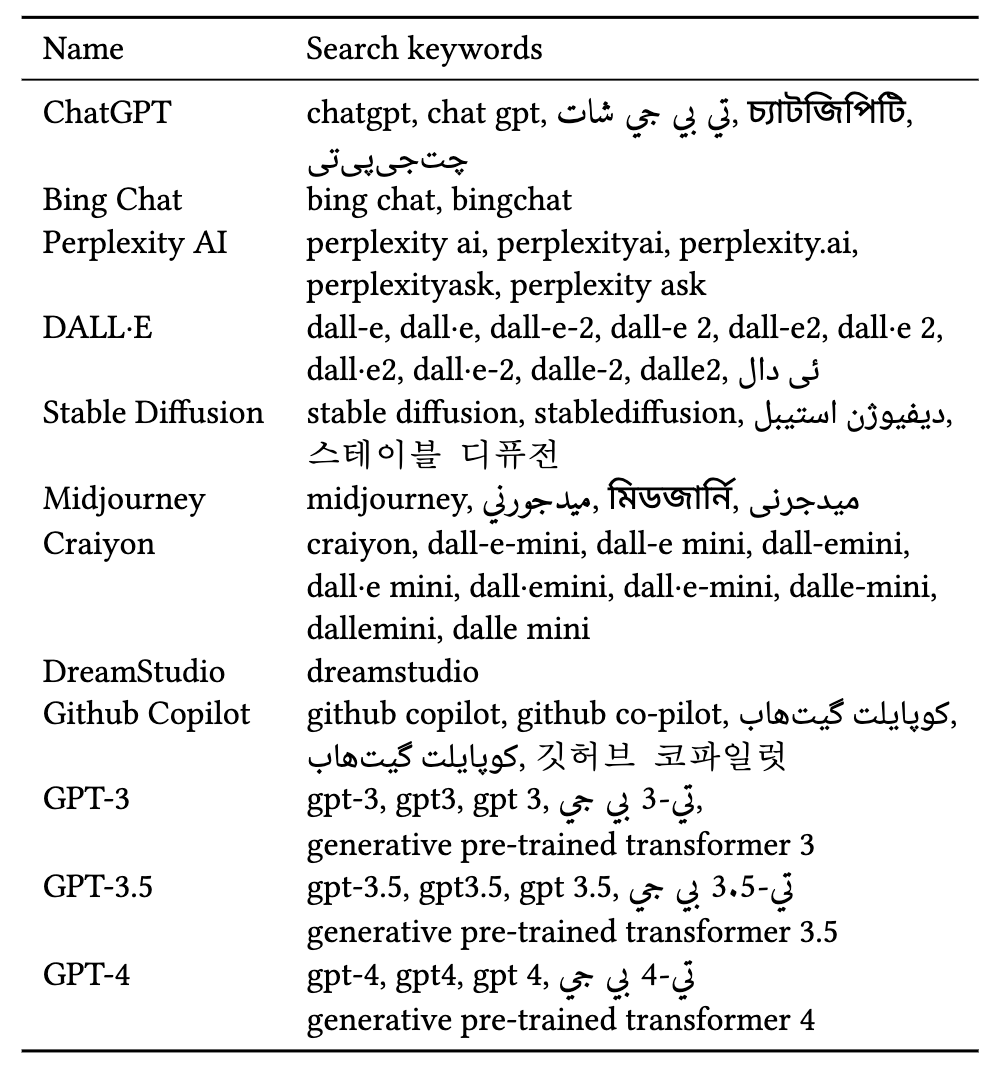}
    \vspace{-1em}
\end{figure}

\section{Search keywords for generative AI tools}\label{ap1}
We aimed to gather a comprehensive collecting of tweets related to generative AI tools utilizing the Twitter Academic API. 
In alignment with Twitter's search conventions, we employed case-insensitive search terms and explored multiple variations, drawing inspiration from \cite{miyazaki2024public}.
Furthermore, recognizing that tool names may vary across languages, we cross-referenced the Wikipedia page for each service in every language. 
From these pages, we extracted the tool name found in the title and incorporated it into our search keywords.
A comprehensive listing of our search keywords is presented in Figure~\ref{search_keyword}.
To ensure that disparities in the representation of tool names across various languages do not introduce biases into our analysis results, we systematically substituted all tool names in the tweets with their accurate and standardized names, as provided in the first column of Figure~\ref{search_keyword}.
We publish the collected Tweet IDs using the Twitter Academic API at \url{https://github.com/hkefka385/ICWSM2025_Perspective_GenerativeAI}.

\section{Details of noise removal and preprocessing}~\label{detail_prepro}

\noindent \textbf{Noise removal.}
To enhance the accuracy and reliability of our subsequent analysis, we implemented a three-step noise removal process on the collected tweets. 
Specifically, we removed tweets that met the following criteria: 1) tweets where the author's name or mentioned usernames include the name of the generative AI tool; 2) tweets discussing a generative AI tool before its release date; and 3) tweets originating from bot-like accounts.
The first noise removal step involved eliminating tweets where the tool name appeared in the author’s username or in mentions, as these were often promotional or irrelevant content, rather than genuine user discussions.
The second step filtered out tweets posted before the official release of a generative AI tool. 
Tweets made before the release date may contain speculation, rumors, or unrelated references, which could introduce noise and compromise the accuracy of our analysis. 
Additionally, the release timeline for different tools varies: some are announced and released simultaneously, while others are announced in advance. 
By excluding tweets made before each tool’s release date, we ensured consistency in the data, allowing for a fair comparison between different tools.
For the third step, which aimed to identify and eliminate bot-like accounts, we employed Botometer~\cite{yang2022botometer}, a widely recognized tool for bot detection. 
Botometer assigns a bot score on a scale from 0 to 1 to Twitter accounts.
To ensure a robust analysis and minimize the influence of active bots, we scrutinized users who had posted more than 10 tweets, amounting to 218,949 users in total. 
We assessed whether these users exhibited bot-like behavior by applying a threshold of 0.43, following a relatively conservative setting, as described ~\cite{rauchfleisch2020false}.
This process led to the identification of 38,349 (17.5\%) users as bots, and we subsequently removed a total of 1,848,744 tweets associated with these bot accounts.

\noindent \textbf{Preprocessing.}
For the retrieved tweet content, we began by tokenizing, lower-casing, and removing non-word characters, as well as each language stopwords based on NLTK~\cite{bird-loper-2004-nltk}. 
We also removed mentions of other users, URLs, and the hashtag sign.
The tool names were replaced with their proper names (see the second column of Table 1), in order to ensure that differences in the notation of tool names in the languages did not affect the analysis results.

\begin{figure}
    \centering
    \includegraphics[width=8.5cm]{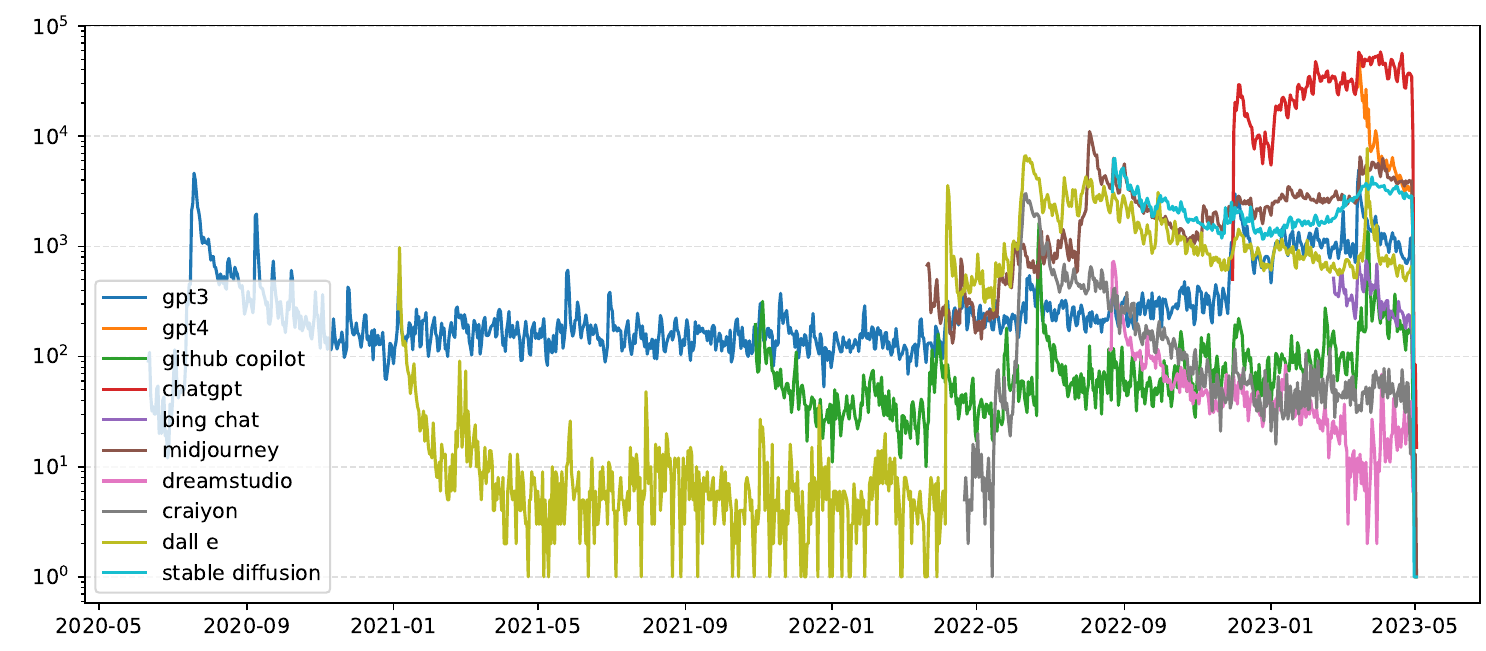}
    \caption{Time series of tweets about generative AI tools}
    \label{time_generative}
\end{figure}

\begin{figure}
    \centering

    \begin{subfigure}[b]{8.5cm}
        \centering
        \includegraphics[width=8.5cm]{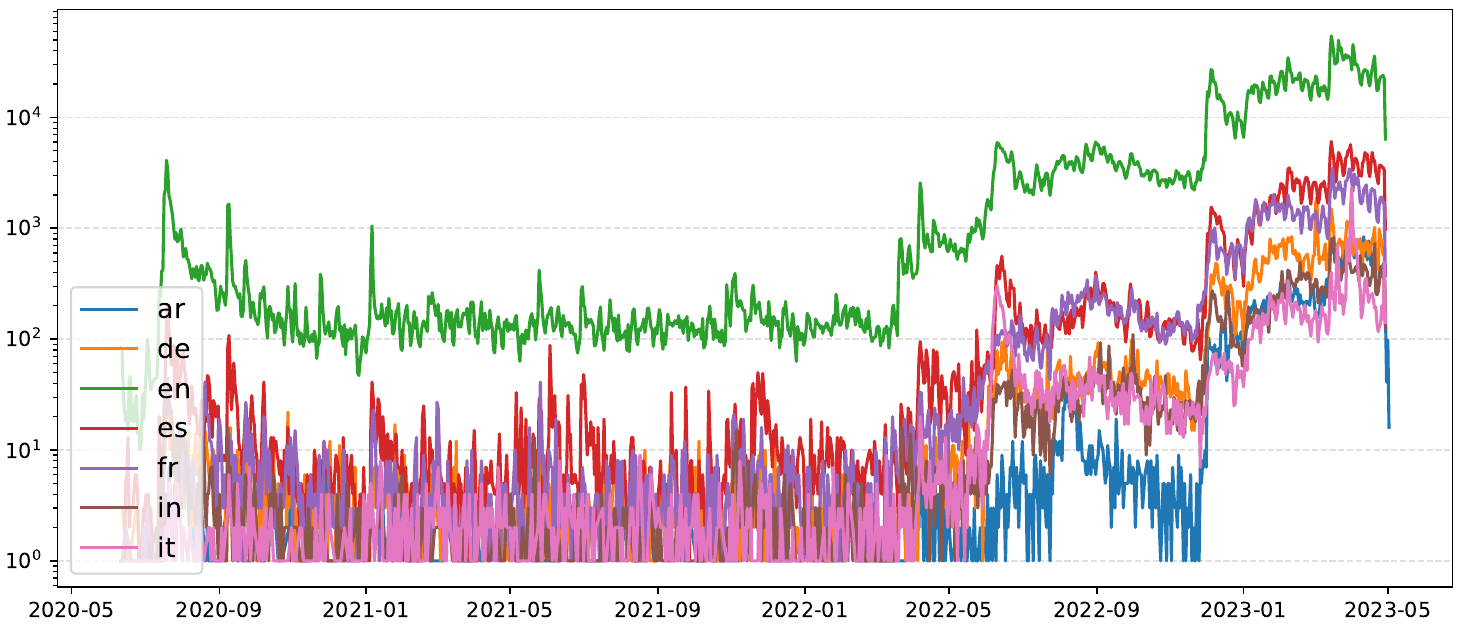}
        \captionsetup{justification=centering, labelsep=period}
        \caption{Time series of tweets about \texttt{ar}, \texttt{de}, \texttt{en}, \texttt{es}, \texttt{fr}, \texttt{in}, and \texttt{it}.}
    \end{subfigure}

    \vspace{0.5cm}
    \begin{subfigure}[b]{8.5cm}
        \centering
        \includegraphics[width=8.5cm]{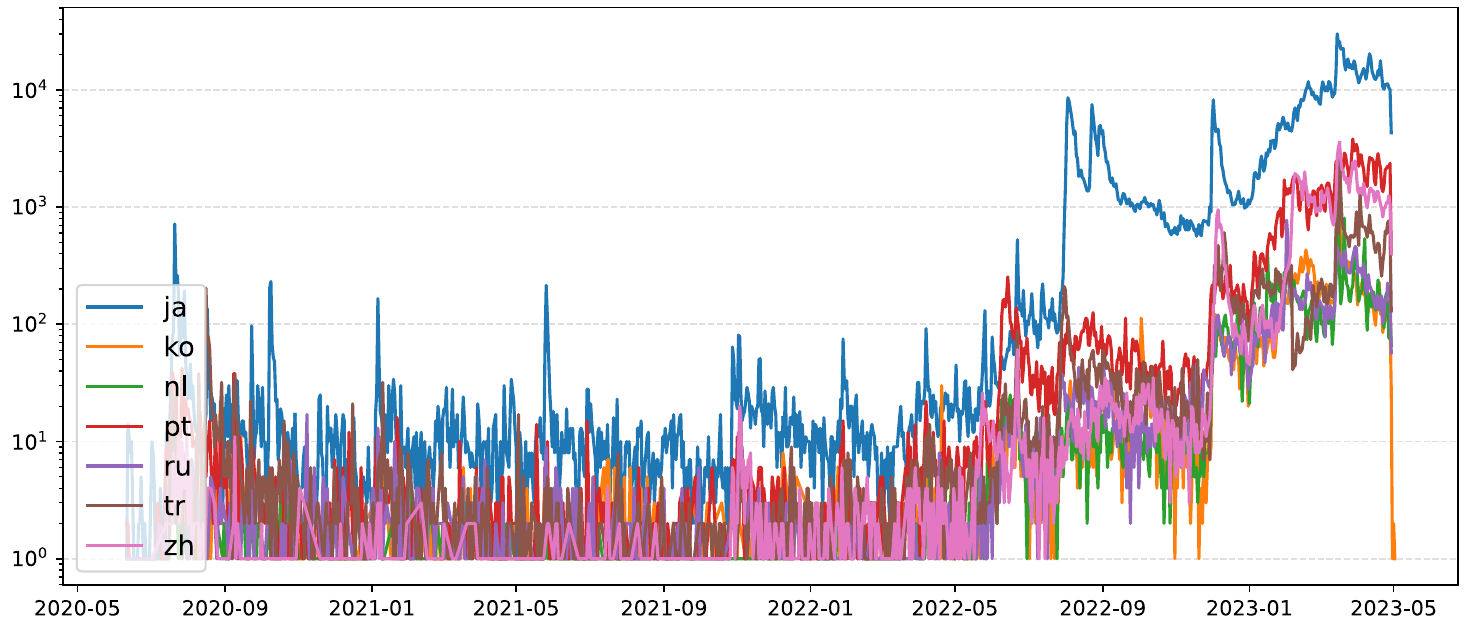}
        \captionsetup{justification=centering, labelsep=period}
        \caption{Time series of tweets about \texttt{ja}, \texttt{ko}, \texttt{nl}, \texttt{pt}, \texttt{ru}, \texttt{tr}, and \texttt{zh}.}
    \end{subfigure}

    \caption{Time series of tweets for each language.}
    \label{time_language}
\end{figure}

\begin{figure}
    \centering
    \includegraphics[width=8.5cm]{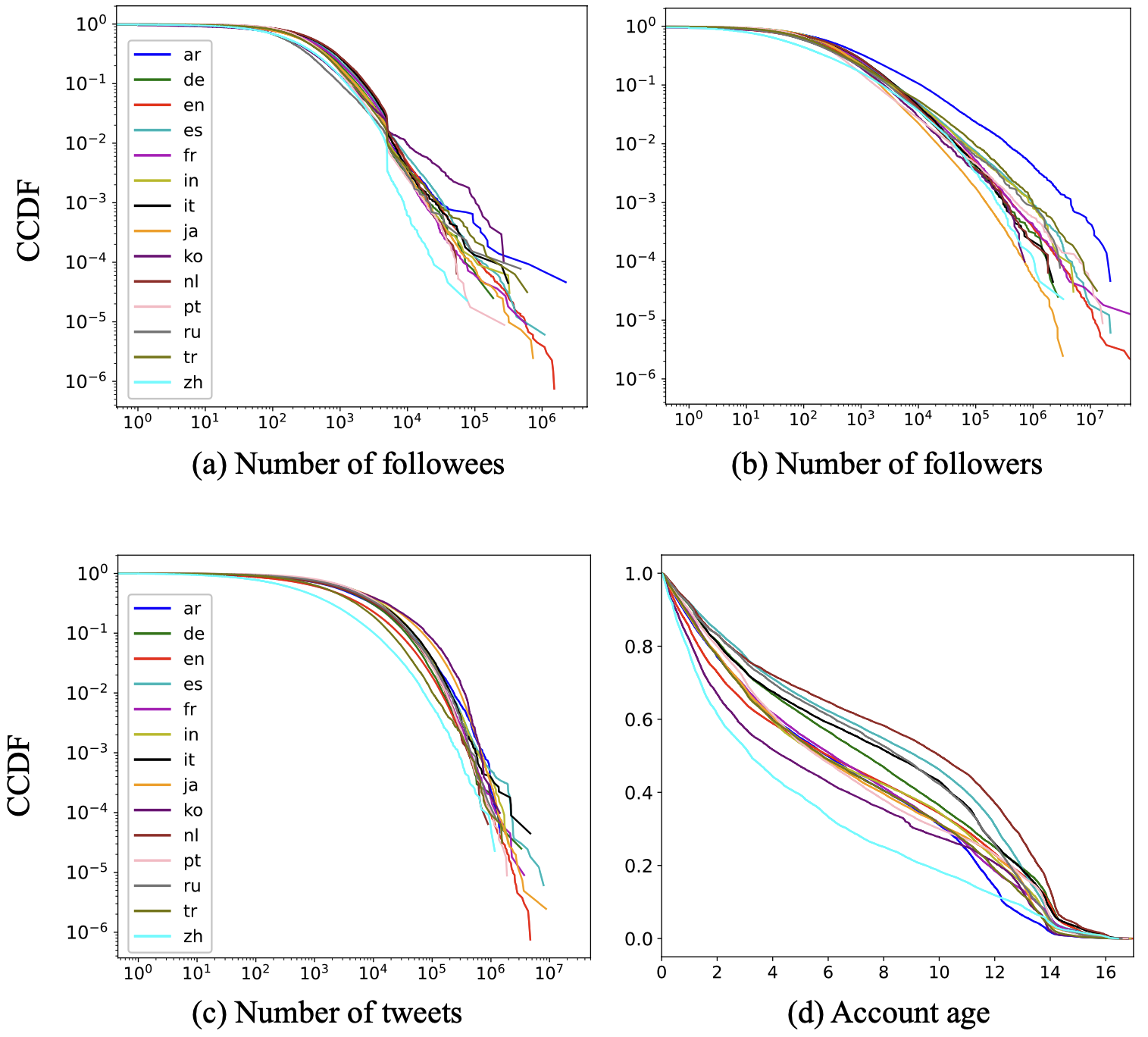}
    \caption{Complementary Cumulative Distribution Functions (CCDF) for User Statistics for each language. 
    Each color represents a user of each language.
    (a) The number of followees, (b) The number of followers, (c) The number of tweets, (d) Account age (years).}
    \label{user_statistics}
\end{figure}

\section{Time series of tweets about generative AI tools}
Figure~\ref{time_generative} presents a time series of tweets for each AI tool.
While the topic of generative AI tools saw modest activity in 2020 and 2021, the release of DALL·E 2 in April 2022 activated discussions around image generation services, including Craiyon and Midjourney. 
Notably, the release of ChatGPT garnered significant public attention: tweets mentioning ChatGPT ranged from 10,000 to 100,000, a striking contrast to the previous peak of 10,000 daily tweets for other tools.

Figure~\ref{time_language} presents a time series of tweets for each language.
Most languages experienced an increase in tweet volume following the release of DALL·E 2, signaling a heightened interest in image generation tools.
In the case of \texttt{ja}, the release of Midjourney spurred a pronounced surge in tweet activity, reflecting a strong engagement with this specific tool.
\texttt{en} shows that tweet activity maintained a steady baseline of approximately 100 tweets per day even prior to the release of DALL·E 2
This suggests that discussions on generative AI tools were already active in the linguistic area before the wider interest grew.

\section{User statistics for each language}
Figure~\ref{user_statistics} illustrates the distributions of various user characteristics, such as the number of followees, followers, tweets, and account ages, across different language communities.
At the number of followers, within the \texttt{ar} community, approximately 2\% of users discussing generative AI tools have over 100,000 followers. 
This is in stark contrast to the \texttt{ja} community, where only approximately 0.2\% of users reach similar follower counts.
This significant disparity suggests a more influencer-centric discourse within the \texttt{ar} community, as opposed to the \texttt{ja} community where generative AI topics are more widely discussed among users with smaller follower bases.
Additionally, the \texttt{zh} community shows a unique trend in account ages, with over 40\% of users having created their accounts subsequent to the advent of generative AI, which means an account age of less than three years.
This percentage is high compared to other language communities.

\begin{table}
  \footnotesize
  \caption{List of sentiment models}
  \label{sentiment_model}
  \scalebox{0.66}{
  \begin{tabular}{clr}
  \toprule
  Language & Model & Accuracy\\ \midrule
  en & cardiffnlp/twitter-roberta-base-sentiment \cite{en_sent} & 0.78 \\
  ja & jarvisx17/japanese-sentiment-analysis \cite{ja_sent} & 0.88 \\
  es & cardiffnlp/xlm-v-base-tweet-sentiment-es~\cite{es_sent} & 0.68 \\
  fr & cardiffnlp/xlm-roberta-base-tweet-sentiment-fr~\cite{fr_sent} & 0.72 \\
  pt & cardiffnlp/xlm-v-base-tweet-sentiment-pt~\cite{pt_sent} & 0.80\\
  zh & SnowNLP~\cite{zh_sent} & 0.88\\
  de & cardiffnlp/xlm-v-base-tweet-sentiment-de~\cite{de_sent} & 0.72\\
  tr & savasy/bert-base-turkish-sentiment-cased~\cite{tr_sent} & 0.84\\
  id & w11wo/indonesian-roberta-base-sentiment-classifier~\cite{id_sent} & 0.74\\
  ar & CAMeL-Lab/bert-base-arabic-camelbert-da-sentiment~\cite{ar_sent}& 0.78\\
  it & osiria/bert-tweet-italian-uncased-sentiment~\cite{it_sent} & 0.84\\
  ru & blanchefort/rubert-base-cased-sentiment-rusentiment~\cite{ru_sent} & 0.66\\
  ko & matthewburke/korean\_sentiment~\cite{ko_sent} & 0.84\\
  nl & DTAI-KULeuven/robbert-v2-dutch-sentiment~\cite{nl_sent} & 0.86\\
  \bottomrule
\end{tabular}
}
\end{table}

\section{List of sentiment classification models}\label{ap2}
Table~\ref{sentiment_model} presents the list of sentiment classification models employed for tweets in each language.
These models classify tweets into either two categories (Positive and Negative) or three categories (Positive, Negative, and Neutral).
To evaluate their performance, we manually assessed the classification of 50 tweets in each language from our collected dataset. 
We found that the accuracy of all models ranged from 0.66 to 0.88, which we deemed adequate for analyzing the vast number of tweets in this study.

\begin{figure*}
    \centering
    \includegraphics[width=18cm]{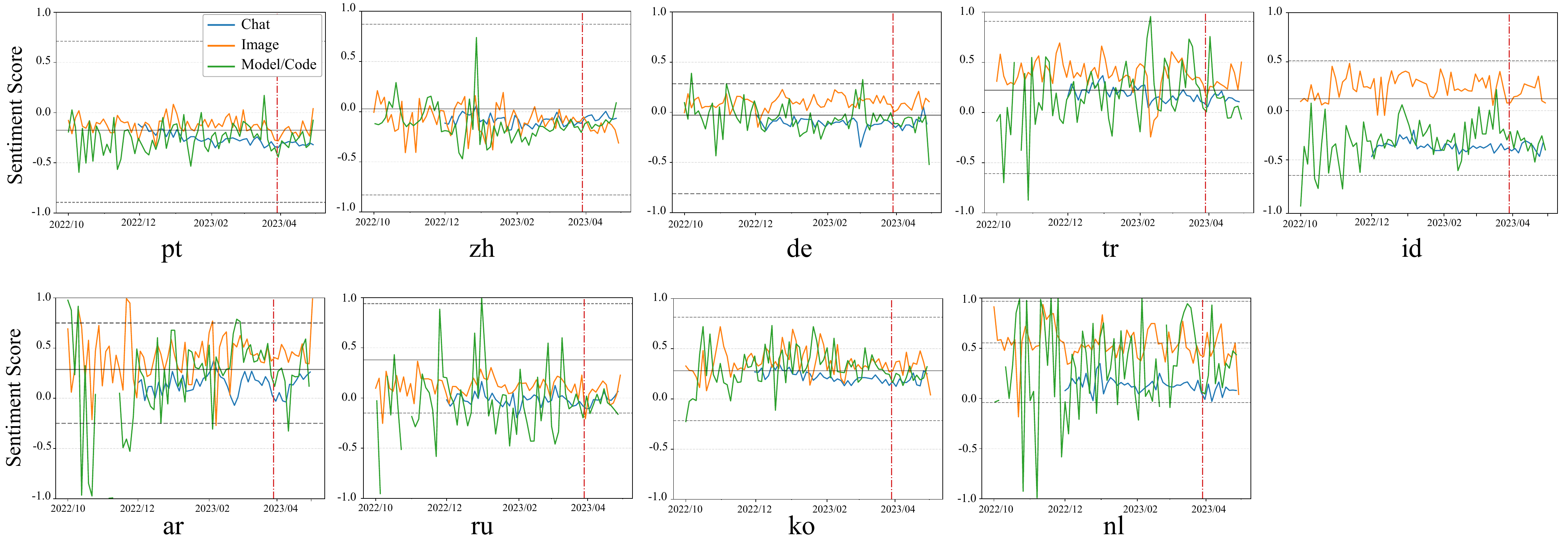}
    \caption{Time series of sentiment score in nine languages: \texttt{pt}, \texttt{zh}, \texttt{de}, \texttt{tr}, \texttt{id}, \texttt{ar}, \texttt{ru}, \texttt{ko}, and \texttt{nl} from October 2022 to May 2, 2023. 
    The x-axis represents time, and the y-axis represents the sentiment score. 
    The sentiment scores for generative AI tools are displayed as follows: the blue line represents Chat tool, orange represents Image tool, and green represents Model/Code tool.
    The solid gray line indicates the daily sentiment score, while the gray dotted lines represent the 75th and 25th percentiles of daily sentiment scores calculated from random tweets.
    The vertical red dotted line marks the date on which Italy banned access to ChatGPT.}
    \label{sentiment_trans_appendix}
\end{figure*}

\begin{figure}
    \centering
    \includegraphics[width=8.5cm]{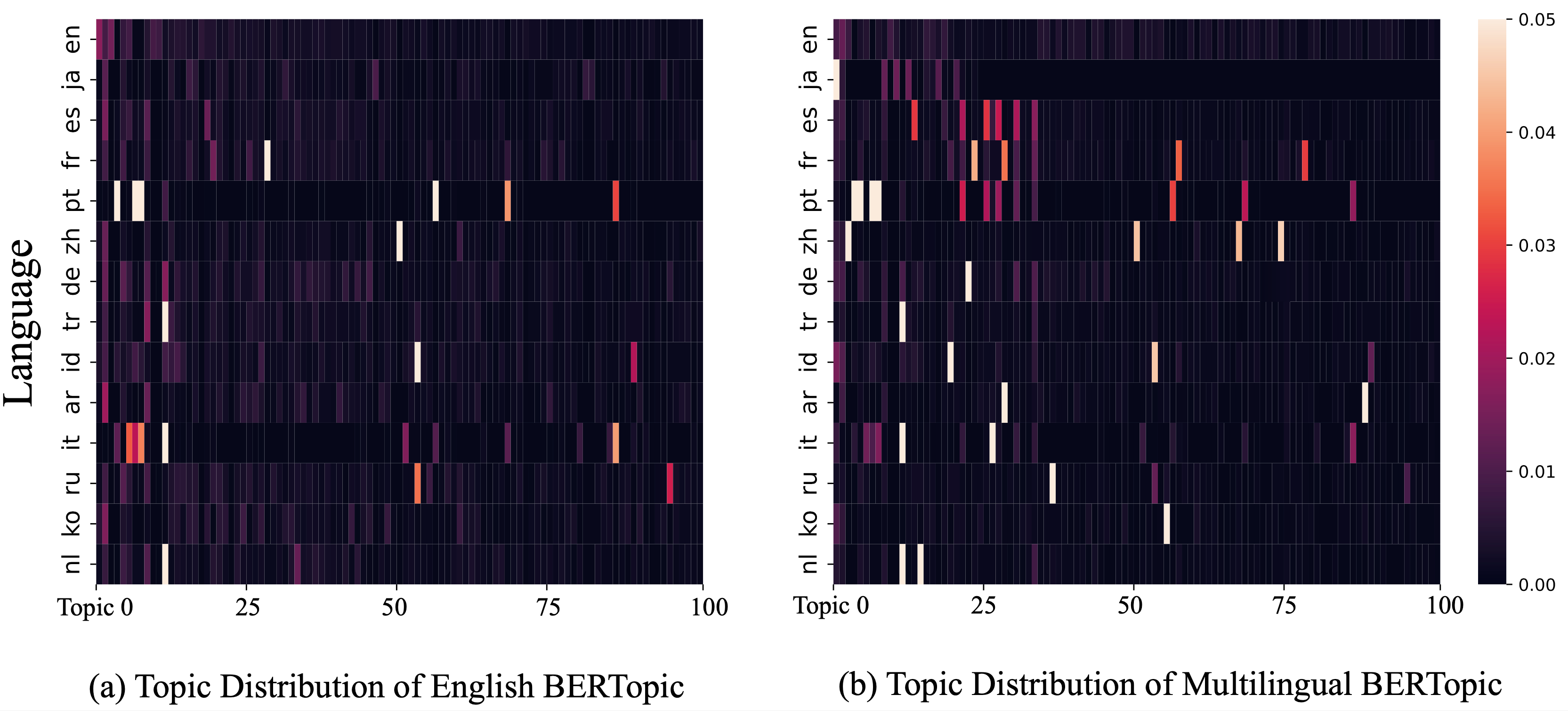}
    \caption{Heatmaps representing the top 100 topic distribution of (a) English BERToopic and (b) Multilingual BERTopic across 14 languages.
    Each column corresponds to a different topic, while each row represents a language. 
    The color intensity indicates the ratio of each topic within each language.
    }
    \label{bertopic}
\end{figure}

\section{Analysis of prominent words in sentiment classification}

To gain insight into the nature of the sentiment toward generative AI tools, we extended our analysis to identify prominent words for both positive and negative sentiment tweets.
Specifically, tweets were categorized into the top 20\% for positive sentiment and the bottom 20\% for negative sentiment, using language-specific sentiment classification models. 
Common stop words and tool names were excluded, and the odds ratio for each word was calculated within these groups to highlight its prominence.

\begin{figure*}
\centering
  \caption{Top 10 words associated with positive and negative sentiment tweets toward generative AI tools across multiple languages, except generative AI tool' names.}
  \label{sentiment_words}
  \includegraphics[width=18cm]{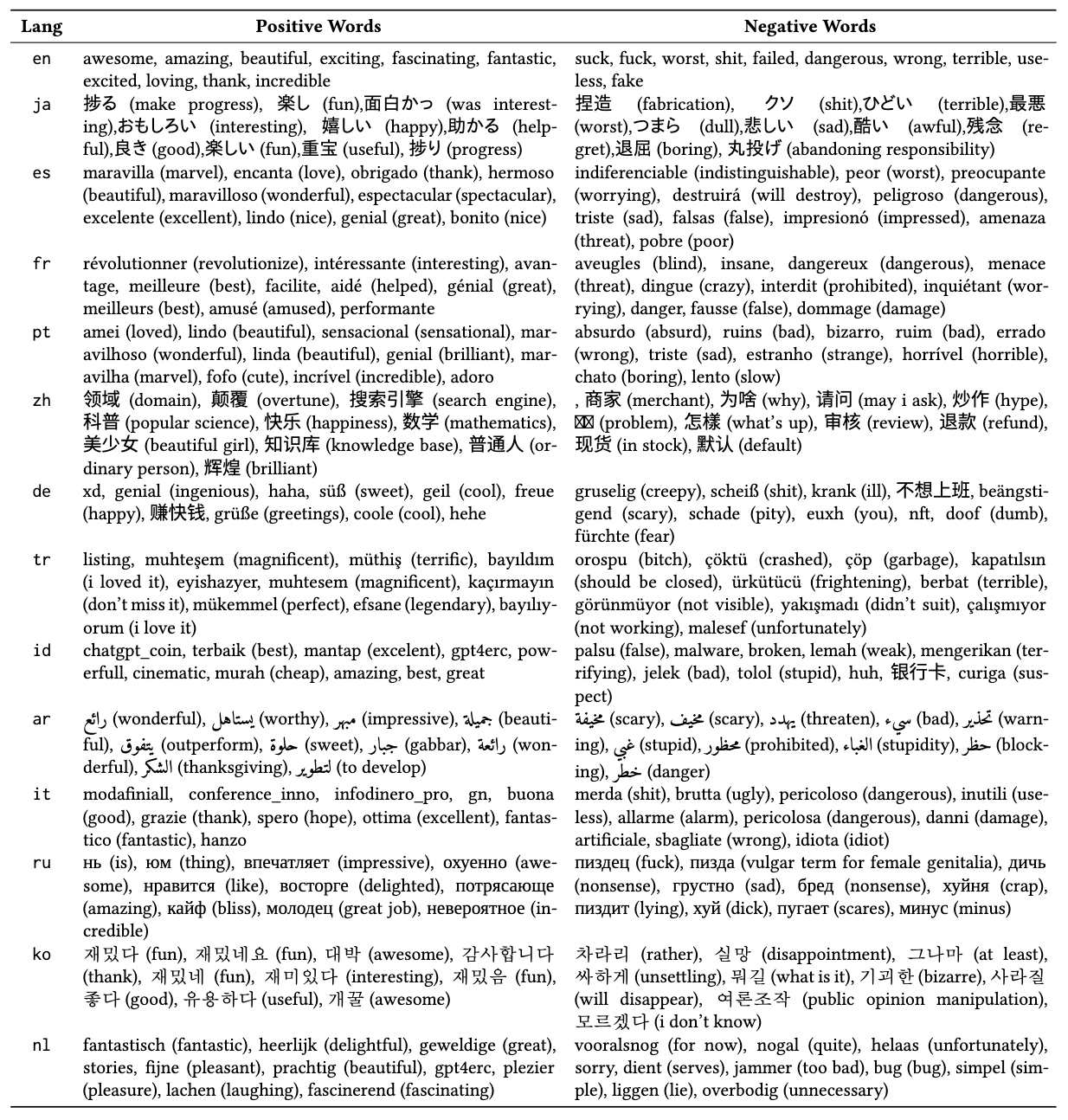}
\end{figure*}

Figure~\ref{sentiment_words} illustrates the top 10 words for different language communities ranked by their odds ratios.
Across all languages, words expressing surprise or amazement were prominently associated with positive sentiment, such as ``interesting,'' ``excellent,'' and ``impressive.''
These words indicate that users are often impressed by the generative AI tools, either when trying them out or reflecting on their functionality.
These words reflect a sense of fascination and positive reactions toward the innovative capabilities of the tools.
Conversely, words associated with negative sentiment often focused on concerns about the accuracy and reliability of the tools' outputs. 
For instance, words like ``fake'' and ``false'' indicate worries about the authenticity or correctness of the outputs, while words such as ``horrible'' and ``danger'' point toward anxiety regarding the potential future impacts or risks of these technologies.
These negative sentiments reflect a cautious or skeptical attitude, particularly regarding the trustworthiness of AI-generated content.

An interesting case is observed in the \texttt{zh} community. 
Unlike other languages, where emotional words dominate both positive and negative sentiment groups, more technical words such as ``domain,'' ``search engine,'' and ``knowledge base'' emerge in the Chinese context.
This suggests that users in this community may focus more on the mechanics and utility of generative AI rather than purely emotional reactions. 
It points to a more pragmatic view, where users are interested in how these tools can be applied within specific domains of expertise or as knowledge resources.

\begin{figure*}
\centering
  \caption{Top 10 words associated with positive and negative sentiment tweets for Chat, Image, and Model/Code tools across four languages with the highest number of tweet (\texttt{en}, \texttt{ja}, \texttt{es}, and \texttt{fr}), excluding generative AI tool names.}
  \label{sentiment_words2}
  \includegraphics[width=18cm]{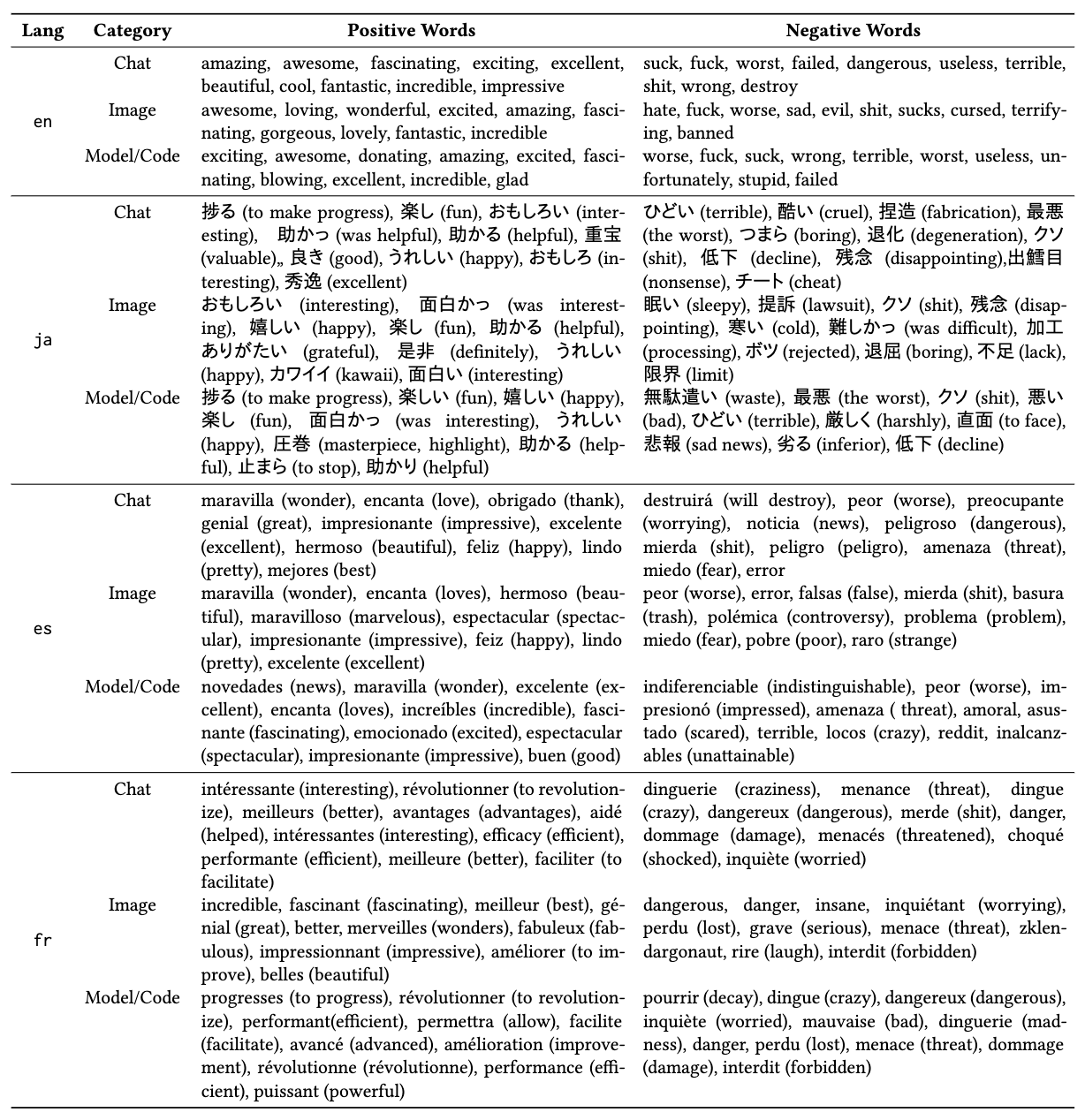}
\end{figure*}

Additionally, by focusing on the four languages with the highest number of tweets (\texttt{en}, \texttt{ja}, \texttt{es}, and \texttt{fr}), Figure~\ref{sentiment_words2} shows the prominent words associated with positive and negative sentiments for each tool category (Chat, Image, Model/Code.) 
In the positive sentiment group for Chat tools, words expressing surprise and fascination, such as ``fantastic'' and ``interesting,'' were common. 
Conversely, for Image tools, positive sentiment tweets frequently included words indicating affection and enjoyment, such as ``love'' and ``beautiful,'' indicating that users feel aesthetic pleasure and emotional satisfaction from the outputs of image generation tools. 
These differences highlight how the type of tool influences the nature of positive emotional responses it elicits.

In contrast, negative sentiment tweets for Chat tools prominently featured words like ``destroy'' and ``danger,'' which are not commonly seen with other tools. 
These words reflects fears of disruption to existing norms and concerns over potential negative impacts on society, such as job displacement or ethical issues related to AI usage. 
These findings support the hypothesis of contrasting sentiment results observed between Image and Chat tools, as discussed in Section 4. 
Emotional reactions to each tool type appear to be influenced by their functionalities and the perceived impact on users' lives.

\begin{table}
  \small
  \caption{Coherence score and of two BERTopic models}
  \label{bertopic_performance}
  \scalebox{0.92}{
  \begin{tabular}{rrr}
    \toprule
    & English BERTopic & Multilingual BERTopic\\ \midrule
    \multicolumn{1}{l}{Coherence (Overall)} & 0.513 & 0.535\\ \midrule
    \multicolumn{1}{l}{Diversity (Overall)} & 0.688& 0.709\\
    \texttt{en} & 0.618 & 0.675\\
    \texttt{ja} & 0.648 & 0.884\\
    \texttt{es} & 0.698 & 0.791\\
    \texttt{fr} & 0.717 & 0.828\\
    \texttt{pt} & 0.857 & 0.843\\
    \texttt{zh} & 0.736 & 0.887\\
    \texttt{de} & 0.763 & 0.853\\
    \texttt{tr} & 0.784 & 0.904\\
    \texttt{id} & 0.800 & 0.854\\
    \texttt{ar} & 0.789 & 0.906\\
    \texttt{it} & 0.889 & 0.891\\
    \texttt{ru} & 0.810 & 0.879\\
    \texttt{ko} & 0.801 & 0.867\\
    \texttt{nl} & 0.867 & 0.896\\
  \bottomrule
\end{tabular}
}
\end{table}

\section{Comparing English BERTopic and Multilingual BERTopic}\label{bertopic_compare}
In analyzing multilingual content, we examine whether employing Multilingual BERTopic based on all-MiniLM-L6-v2~\cite{multilingual_sentence} is more effective, or if translating the content into English and applying Monolingual BERTopic (English BERTopic) based on \cite{grootendorst2022bertopic} is preferable. 
The analysis is based on the dataset detailed in Section 5.2. 
When applied to this dataset, English BERTopic identified 4,420 topics, while Multilingual BERTopic yielded 4,501 topics, showing no significant difference in the number of topics generated.

Table~\ref{bertopic_performance} outlines Coherence and Topic Diversity scores for both models.
Coherence in a topic model reflects its interpretability, where a higher coherence score denotes a model's enhanced capacity to generate consistent and meaningful topics.
In this aspect, Multilingual BERTopic outperformed its English BERTopic. 
And, Topic Diversity measures the variety and distinctness of the generated topics. 
Multilingual BERTopic demonstrates superior topic diversity across various languages, suggesting a more nuanced and language-specific topic generation.

Figure~\ref{bertopic} illustrates the distribution of topics assigned by the two BERTopic models. 
The distribution of topics in English BERTopic tends to assign a greater variety of topics than in Multilingual BERTopic. 
On the other hand, Multilingual BERTopic exhibited a higher occurrence of language-specific topics. 
Also, the higher similarity in topics between \texttt{es} and \texttt{pt} implies that linguistic similarities significantly influence topic generation.
These observations indicate that Multilingual BERTopic's proficiency in creating detailed, language-tailored topics.

Considering the goal of extracting common topics across languages, we opted for English BERTopic in our analysis.

\section{Visualization of topic networks by users who posted multiple times}

We aim to uncover the connections between topics by analyzing the behavior of users who engage with multiple topics. 
This builds on the findings from RQ2, where we identified key topics within different linguistic communities and observed common topics across multiple language communities. 
While RQ2 insights into the prominence of specific topics, it did not fully explore how these topics are interconnected through user activity.
To address this gap, we specifically focused on users who posted about more than one topic.
When a user posts with multiple topics about generative AI tools, we infer that there is a meaningful relationship between those topics.
For visualization of the relationship, we present network of topics, where each topic is represented as a node and the edges between nodes indicate that a user has posted about both topics.
This visualization reveal how certain topics cluster together, suggesting broader thematic relationships.

\begin{figure*}
    \centering
    \includegraphics[width=18cm]{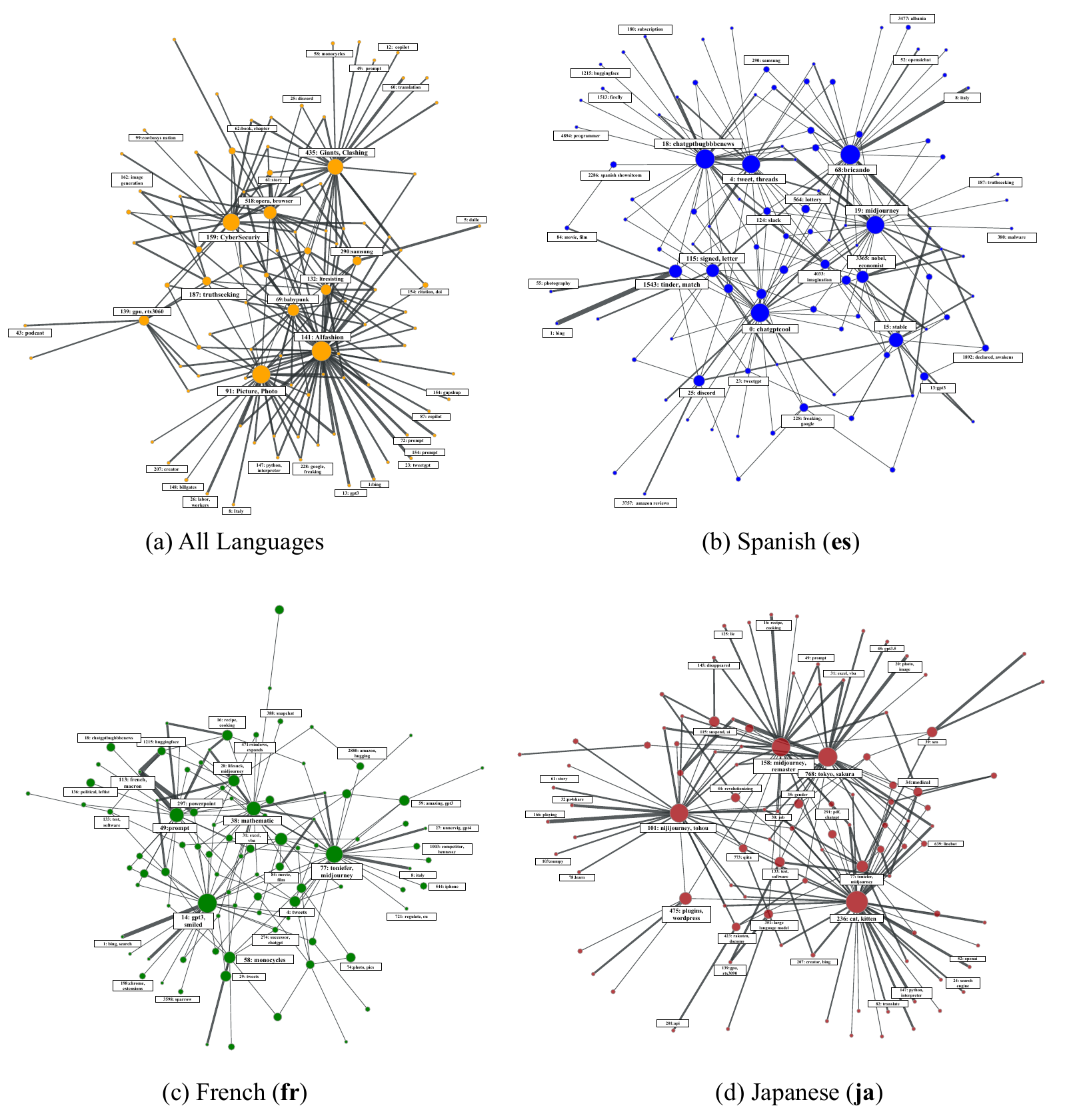}
    \caption{Network of topics linked by user activity across different languages. 
    The size of each node represents the number of topics that users who posted multiple topics have contributed to. 
    The thickness of the edges between nodes indicates the frequency with which each user posted about both the connected topics.}
    \label{topic_visualize}
\end{figure*}

Figure~\ref{topic_visualize} shows the visualization of major topics for (a) all languages, (b) Spanish (\texttt{es}), (c) French (\texttt{fr}), and (d) Japanese (\texttt{ja}).
Although lower topic numbers in RQ2 corresponded to higher tweet counts, the focus on users who posted multiple times about generative AI tools shows that frequently discussed topics in this experimental setting differ from the generally popular ones.
For instance, across all languages, topics like 141: AIfashion and 91: Picture, related to image generation, are prevalent among users who actively engage with these tools. 
Additionally, these users show strong interest in programming language (147: Python) and input methods for generative AI (72,154: Prompt).

When examining individual languages, unique linguistic trends emerge.
In \texttt{fr}, the French national topic (113: French, Macron) is closely linked to technical subjects such as 1215: Huggingface, 49: Prompt, and 297: Powerpoint, indicating that users involved in political discussions also have a strong interest in technology.
Similarly, in \texttt{ja}, the network reveals connections between 101: nijijourney (anime image generation) and programming-related topics like 103: numpy and 78: learn, suggesting that users generating anime-style images are also experimenting with Python programming.
Interestingly, topic 236: cat highlights a distinct focus on cat image generation, which is notably specific to the Japanese-speaking community.

\section{Details of extraction interactions from images}\label{ap_extraction}
We analyzed screenshots of interactions with ChatGPT, which many users voluntarily share on Twitter.
The interaction extraction consists of the following two steps: 1) Rule-based classification of ChatGPT screenshots; and 2) Conversion of the extracted images into text using optical character recognition (OCR).

\begin{figure}
\centering
  \begin{subfigure}[t]{.47\linewidth}
    \centering
    \includegraphics[width=\linewidth]{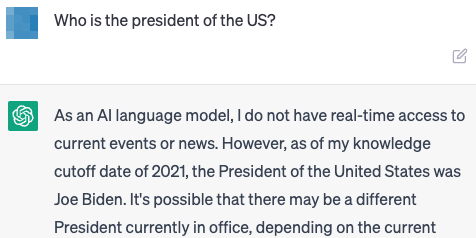}
    \caption{Light mode.}
    \label{fig:1a_sb}
  \end{subfigure}
  \begin{subfigure}[t]{.47\linewidth}
    \centering
    \includegraphics[width=\linewidth]{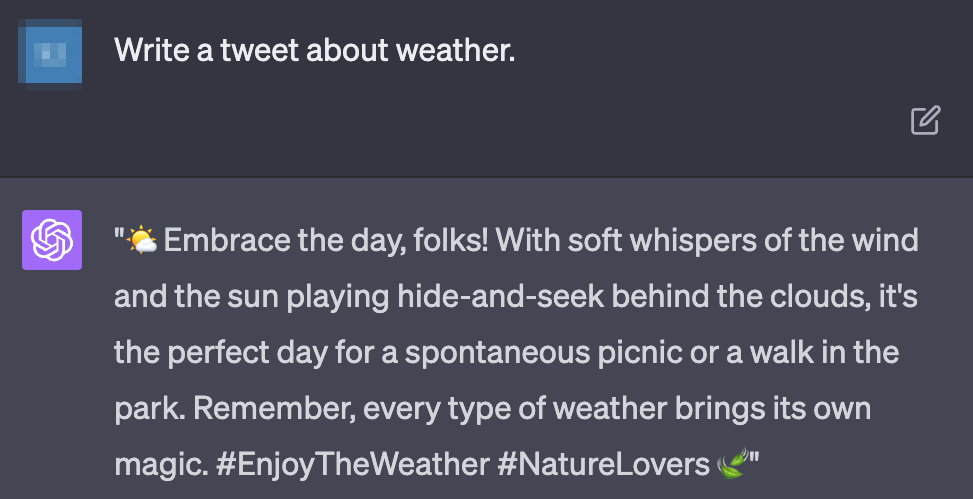}
    \caption{Dark mode.}
    \label{fig:1b_sb}
  \end{subfigure}
  \setlength{\abovecaptionskip}{4pt plus 3pt minus 2pt}
  \caption{Examples of ChatGPT images. The upper part is the prompt section, and the bottom is the response section in both images.}
  \label{chatgpt_images} 
\end{figure}

\noindent \textbf{1) Rule-based classification of ChatGPT screenshots: }
The design of ChatGPT interaction images typically incorporates a single background color that clearly distinguishes the prompt and response sections, featuring two primary themes: light mode and dark mode, as depicted in Figure~\ref{chatgpt_images}. 
This distinct design facilitates the application of rule-based methods for image extraction.
To identify an image as a ChatGPT screenshot, we verified whether two out of the top three RGB colors in the image fall within the RGB range of the prompt and response sections.
In light mode, the subpixel RGB values for the prompt section should fall within the range $(R,G,B)\in[251,255]\times[251,255]\times[251,255]$, while for the response section, they should be within  $(R,G,B)\in[251,255]\times[251,255]\times[251,255]$.
In dark mode, the RGB values for the prompt section should be in the range $(R,G,B)\in[51,54]\times[51,54]\times[63,66]$, and for the response section, they should be within $(R,G,B)\in[67,71]\times[67,71]\times[81,85]$. 
This straightforward approach achieved an F1 score of 0.90 in testing on 200 randomly selected and manually annotated images, demonstrating sufficient accuracy for our analysis.

\noindent \textbf{2) Conversion of the extracted images into text using OCR: }
We employed OCR with the Cloud Vision API~\cite{cloud_vision} to extract text from the prompts and ChatGPT responses identified in the screenshots. 
This API, provided by Google Cloud, supports multiple languages, making it suitable for handling multilingual texts within the scope of this study.

Through these steps, we successfully retrieved a total of 507,714 interactions with ChatGPT, called as ``Interaction dataset.''
Table~\ref{interaction_statistics} shows the number of interactions and the percentage of tweets about ChatGPT including interactions among ChatGPT tweets in each language.
The percentages across many languages hover around 8\%. 
However, for languages primarily spoken in East Asian countries, specifically \texttt{zh}, \texttt{ja}, and \texttt{ko}, the percentage exceeds 10\%. 
This suggests that there is a prevalent culture in these regions of introducing and discussing how to use ChatGPT.

\begin{table}
  \small
  \caption{Number of interactions in Interaction dataset. Pct. indicates that the percentage of tweets about Chatgpt that are related to interactions.}
  \label{interaction_statistics}
  \begin{tabular}{rrrr}
    \toprule
    Language & Number of interactions & IntI & Pct. \\ \midrule
    English (\texttt{en}) & 207,116 & 1.00 & 8.52\%\\ 
    Japanese (\texttt{ja}) & 135,405 & 0.83 & 13.97\%\\  
    Spanish (\texttt{es}) & 22,382 & 0.46 & 7.39\%\\ 
    French (\texttt{fr}) & 13,325 & 1.12 & 8.08\%\\  
    Portuguese (\texttt{pt}) & 10,800 & 0.25 & 6.51\%\\ 
    Chinese (\texttt{zh}) & 11,971 & 2.91 & 10.91\%\\ 
    German (\texttt{de}) & 7,029 & 1.40 & 9.70\%\\ 
    Turkish (\texttt{tr}) & 3,737 & 0.22 & 9.85\%\\ 
    Indonesian (\texttt{id}) & 3,060 & 0.13 & 8.67\%\\ 
    Arabic (\texttt{ar}) & 3,464 & 0.23 & 8.39\%\\ 
    Italian (\texttt{it}) & 2,054 & 0.39 & 8.23\%\\ 
     Russian (\texttt{ru}) &  1,783 & 0.57 & 9.62\%\\ 
    Korean (\texttt{ko}) & 2,602 & 0.13 & 13.11\%\\  
    Dutch (\texttt{nl}) & 1,507 & 0.58 & 7.99\%\\  
  \bottomrule
\end{tabular}
\end{table}

\begin{table*}
    \footnotesize
    \centering
    \caption{Taxonomy of main categories for chatbot usage.}
    \begin{tabular}{l p{4.5cm} p{8.6cm} p{2cm}} 
    \toprule
        Number & Category Name & Description & Num (Pct.)\\ \midrule
        \textbf{1} & \textbf{Search} & Utilizing chatbots to retrieve specific information akin to web search & 2,335 (35.6\%)\\ 
        \multicolumn{1}{r}{1-1} & Engineering \& Coding & Seeking solutions, code snippets, or explanations related to engineering or coding & 435 (6.6\%)\\
        \multicolumn{1}{r}{1-2} & Political \& Economic & Inquiring about political or economic theories, events, or policies & 183 (2.7\%)\\
        \multicolumn{1}{r}{1-3} & Science \& Technology & Seeking explanations or information regarding science or technology & 382 (5.8\%)\\
        \midrule
        \textbf{2} & \textbf{Questions or Requests Reflecting the Responder's Preferences} & Engaging chatbots to obtain personalized or subjective responses & 1,727 (26.4\%)\\ 
        \multicolumn{1}{r}{2-1} & Recommendation & Requesting suggestions based on preferences or criteria & 170 (2.6\%)\\
        \multicolumn{1}{r}{2-2} & Expert Insight & Seeking specialized knowledge or expert opinions & 746 (11.4\%)\\
        \multicolumn{1}{r}{2-3} & Request for Examples & Asking for specific examples, illustrations, or demonstrations & 153 (2.3\%)\\
        \multicolumn{1}{r}{2-4} & Seeking Advice & Soliciting guidance, counsel, or personal insights & 258 (3.9\%)\\
        \midrule
        \textbf{3} & \textbf{Support for Business or Creative Tasks} & Leveraging chatbots to provide support in business or creative tasks & 1,440 (22.0\%)\\
        \multicolumn{1}{r}{3-1} & Business Support & Obtaining assistance with business-related tasks, analysis, or decisions & 512 (7.8\%)\\
        \multicolumn{1}{r}{3-2} & Creative Support & Seeking inspiration, feedback, or assistance in creative projects & 643 (9.8\%)\\
        \midrule
        \textbf{4} & \textbf{Dialogue} & Engaging in conversational interactions with chatbots & 414 (6.3\%)\\
        \multicolumn{1}{r}{4-1} & Life Counseling & Seeking advice on or discussing personal life situations& 37 (0.6\%)\\
        \multicolumn{1}{r}{4-2} & Role-play Conversations & Participating in imaginative or scenario-based dialogues& 129 (2.0\%)\\
        \midrule
        \textbf{5} & \textbf{Humor, Wit, Riddles} & Engaging with chatbots for amusement, humor, or intellectual challenge & 319 (4.9\%)\\
        \midrule
        \textbf{6} & \textbf{Other} & Various other interactions or inquiries that do not fall under the above categories & 318 (4.9\%)\\
        \multicolumn{1}{r}{6-1} & Inquiries about Chatbot's System & Asking about the technical or functional aspects of chatbot & 132 (2.0\%)\\
        \multicolumn{1}{r}{6-2} & Inquiries about Chatbot's Thinking & Asking about the chatbot’s thinking & 147 (2.2\%)\\
    \bottomrule
    \end{tabular}
    \label{taxonomy_chat_all}
\end{table*}

\section{Examples and descriptions for each category based on the taxonomy of chatbot usage}\label{examples_taxonomy}
We identified six main categories and 13 subcategories of chatbot usage through open coding.
Examples and detailed descriptions are presented below. (A summary of these categories is presented in Table~\ref{taxonomy_chat_all}.)

\noindent \textbf{1. Search:} 
This category refers to the use of chatbots to retrieve specific information, similar to web search engines. 
Chatbots excel at retrieving precise information and interpreting complex terms to provide accurate and reliable responses to user queries.

\noindent Example prompt: \textit{What are the five highest rated films produced by Netflix in 2020?}

\noindent \textbullet \textbf{1-1. Engineering \& Coding:} This subcategory, which is a subcategory of the main Category 1, refers to code snippets and descriptions related to engineering and coding areas.
Inquiries about code composition are common in this subcategory.

\noindent Example prompt: \textit{Please provide an example description of checking for the existence of the JSON request key 'email' using Flask-pydantic.}

\noindent \textbullet \textbf{1-2. Political \& Economic:} This subcategory includes prompts related to political and economic theories, events, and policies. 

\noindent Example prompt: \textit{What is the list of countries in Ukraine that are funding the Institute of Biological Sciences?}

\noindent \textbullet \textbf{1-3. Science \& Technology:} This subcategory includes prompts seeking explanations and information on a broad range of scientific and technical concepts.

\noindent Example prompt: \textit{What is the relationship between population variance and unbiased variance?}

\noindent \textbf{2. Questions or Requests Reflecting the Responder's Preferences:}
This main category refers to personalized or subjective responses based on the user's preferences or the context provided. 
Unlike the ``1. Search'' main category, the focus here is on generating responses that may not have a definite answer, but are customized to the individual's criteria, preferences, or circumstances.

\noindent Example prompt: \textit{What's a good place to have a romantic dinner in Paris?}

\noindent \textbullet \textbf{2-1. Recommendation:} 
This subcategory applies when using a chatbot as a recommendation model, suggesting books, restaurants, movies, etc.

\noindent Example prompt: \textit{Based on my liking for historical fiction, can you suggest a few books?}

\noindent \textbullet \textbf{2-2. Expert Insight:} 
This subcategory requires the chatbot to provide thoughtful answers to questions that don't have a clear solution based on domain expertise.

\noindent Example prompt: \textit{What are the best practices for reducing memory leakage in Java?}

\noindent \textbullet \textbf{2-3. Request for Examples:} 
Users often require examples to better understand a concept or scenario. 
This subcategory includes chatbots that provide illustrative examples, demonstrations, or explanations to fulfill user requests.

\noindent Example prompt: \textit{Can you provide an example of a balanced chemical equation?}

\noindent \textbullet \textbf{2-4. Seeking Advice:} 
This subcategory pertains to seeking advice, counsel, or personal perspectives on various subjects.

\noindent Example prompt: \textit{If Coinable were to launch an L2 for Ethereum what should they call it.}

\noindent \textbf{3. Support for Business or Creative Tasks:}
This main category pertains to the use of chatbots to assist users with various business or creative tasks, promote innovation, and address problems. It demonstrates the valuable contribution that chatbots can make to our work, hobbies, and other areas of our lives.


\noindent \textbullet \textbf{3-1. Business Support:}
This category covers the various forms of business assistance that chatbots can offer. 
They can aid in data analysis, summarizing data, conducting market research, and providing insights on operational optimization.

\noindent Example prompt: \textit{Please summarize the text of this URL.}

\noindent \textbullet \textbf{3-2. Creative Support:}
This subcategory contains prompts related to creative assistance. 
Generating lyrics and generating stories are common types of prompts in this subcategory.

\noindent Example prompt: \textit{Write a song about GG allin being a time-traveling vampire lizard.}

\noindent \textbf{4. Dialogue:}
This main category includes prompts that chatbots are considered a conversation partners for conversation rather than information search.

\noindent Example prompt: \textit{Hey! How are you?}

\noindent \textbullet \textbf{4-1. Life Counseling:}
This subcategory applies when consulting a chatbot for stress, relationship or personal issues. 
The chatbot acts as a non-judgmental sounding board for confidential conversations.

\noindent Example prompt: \textit{How can I keep her company?}

\noindent \textbullet \textbf{4-2. Role-play Conversations:}
This category applies to prompts for role-play conversations between users and chatbots.
The user transforms the chatbot into an ideal interlocutor by providing the chatbot with the settings of an ideal interlocutor as a prompt.

\noindent Example prompt: \textit{You are HAL9000. Please answer my questions as HAL9000 in the following prompts.}

\noindent \textbf{5. Humor, Wit, Riddles:}
This main category contains prompts for generating humorous content, such as riddles or jokes.

\noindent Example prompt: \textit{Describe the relationship between Italy and France in the form of a joke.}

\noindent \textbf{6. Others:}
The main category includes various other interactions or inquiries not belonging to other categories.

\noindent \textbullet \textbf{6-1. Inquiries about Chatbot's System:}
This subcategory expects users to ask questions about the technical and functional aspects of the chatbot, including its operations, limitations, and developmental framework.

\noindent Example prompt: \textit{What kind of algorithms do you use to understand and respond to questions?}

\noindent \textbullet \textbf{6-2. Inquiries about Chatbot's Thinking:}
This subcategory includes prompts to understand chatbot's own thinking.

\noindent Example prompt: \textit{When do you feel the most pleasure?}

\begin{figure*}
    \centering
    \includegraphics[width=18cm]{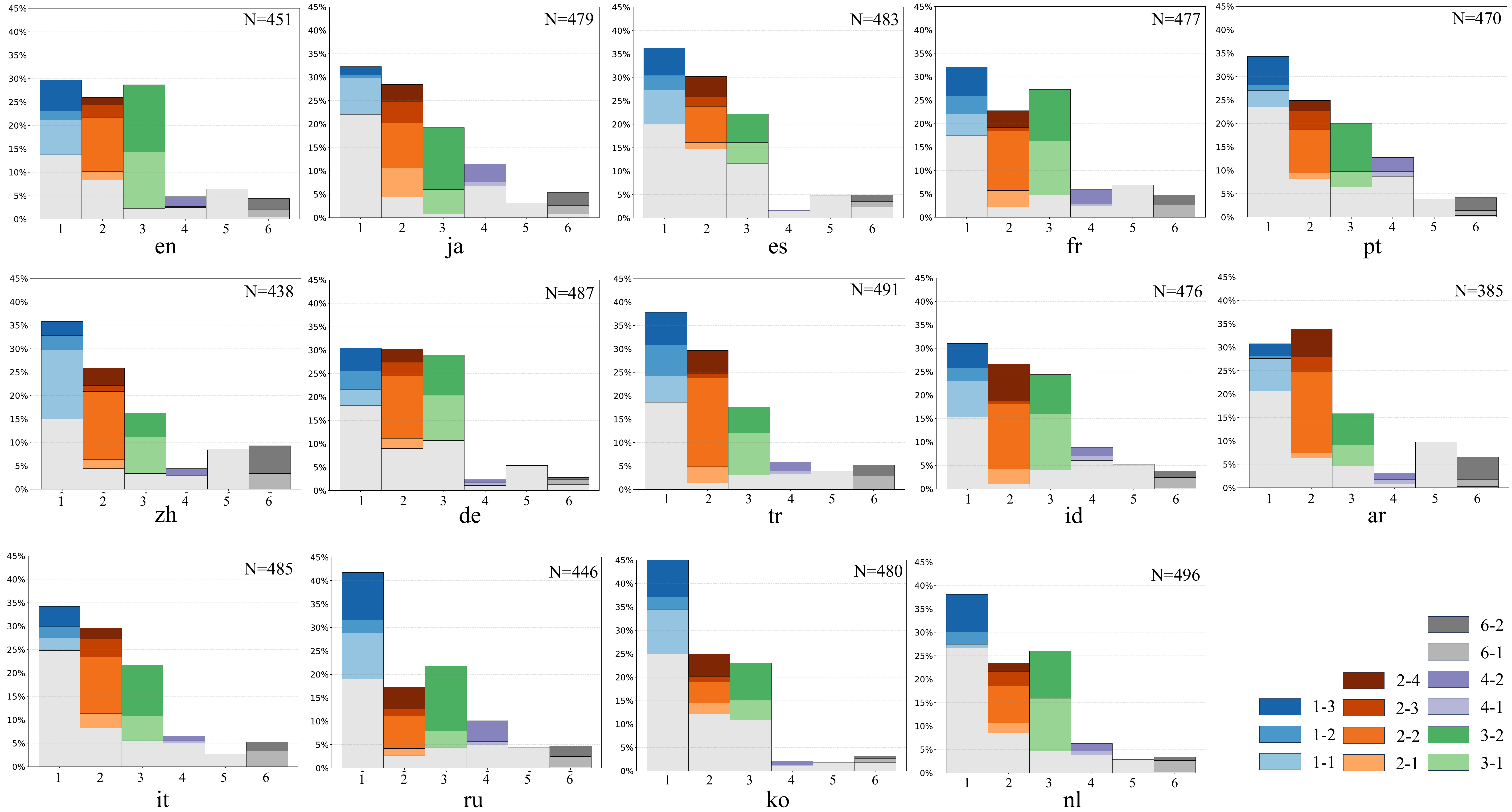}
    \caption{Label distribution for the chatbot usage dataset}
    \label{label_statistics}
\end{figure*}

\section{Rater credibility and expertise}
We identified six main categories and 13 subcategories pertaining to chatbot usage.
Using this taxonomy, the authors formulated annotation guidelines, available at \url{https://github.com/hkefka385/ICWSM2025_Perspective_GenerativeAI}.
Subsequently, a set of 6,553 samples was annotated independently by five annotators.
Of the five annotators, two were authors of this paper who possess expertise and research experience in social media analysis. 
The remaining annotators were Master's and Doctor's students in data engineering who were trained by the authors prior to completing the annotation task.

\begin{table*}
    \small
    \centering
    \caption{Main category classification performance; This reports P: precision, R: recall, F1-score of positive class on the utilization for each category and Macro and Weighted Avg for all samples.}
    \begin{tabular}{lllllllllllll} \toprule
          & \multicolumn{3}{c}{TF-IDF + LR} & \multicolumn{3}{c}{GPT-3} & \multicolumn{3}{c}{BERT} & \multicolumn{3}{c}{DeBERTa}\\
         Variable & P & R & F1 & P & R & F1 & P & R & F1 & P & R & F1 \\
         \midrule
         1 Search & 0.54 & 0.72 & 0.62 & 0.69 & 0.68 & 0.68 & 0.66 & 0.70 & 0.68 & 0.71 & 0.71 & 0.71\\
         \multicolumn{1}{r}{1-1} & 0.68 & 0.75 & 0.71 & 0.74 & 0.86 & 0.79 & 0.76 & 0.80 & 0.78 & 0.68 & 0.68 & 0.68\\
         \multicolumn{1}{r}{1-2} & 0.52 & 0.34 & 0.41 & 0.47 & 0.46 & 0.46 & 0.43 & 0.43 & 0.43 & 0.77 & 0.76 & 0.76\\
         \multicolumn{1}{r}{1-3} & 0.46 & 0.52 & 0.49 & 0.50 & 0.53 & 0.52 & 0.44 & 0.49 & 0.46 & 0.46 & 0.51 & 0.48\\
         2 Preference & 0.35 & 0.24 & 0.29 & 0.39 & 0.35 & 0.37 & 0.39 & 0.32 & 0.35 & 0.41 & 0.32 & 0.37\\
         \multicolumn{1}{r}{2-1} & 0.24 & 0.41 & 0.30 & 0.56 & 0.52 & 0.54 & 0.58 & 0.43 & 0.49 & 0.51 & 0.47 & 0.49\\
         \multicolumn{1}{r}{2-2} & 0.41 & 0.48 & 0.45 & 0.47 & 0.50 & 0.49 & 0.43 & 0.46 & 0.44 & 0.52 & 0.45 & 0.48\\
         \multicolumn{1}{r}{2-3} & 0.32 & 0.45 & 0.38 & 0.45 & 0.45 & 0.45 & 0.50 & 0.19 & 0.28 & 0.50 & 0.51 & 0.50\\
         \multicolumn{1}{r}{2-4} & 0.23 & 0.35 & 0.29 & 0.29 & 0.35 & 0.31 & 0.23 & 0.33 & 0.27 & 0.29 & 0.40 & 0.35\\
         3 Support & 0.30 & 0.31 & 0.30 & 0.24 & 0.27 & 0.26 & 0.45 & 0.44 & 0.45 & 0.49 & 0.47 & 0.48\\
         \multicolumn{1}{r}{3-1} & 0.45 & 0.50 & 0.47 & 0.54 & 0.63 & 0.58 & 0.44 & 0.62 & 0.51 & 0.62 & 0.57 & 0.60\\
         \multicolumn{1}{r}{3-2} & 0.58 & 0.58 & 0.58 & 0.70 & 0.72 & 0.71 & 0.73 & 0.73 & 0.73 & 0.75 & 0.77 & 0.76\\
         4 Dialogue & 0.45 & 0.50 & 0.47 & 0.36 & 0.34 & 0.35 & 0.33 & 0.34 & 0.33 & 0.38 & 0.33 & 0.35\\
         \multicolumn{1}{r}{4-1} & 0.20 & 0.14 & 0.17 & 0.00 & 0.00 & 0.00 & 0.00 & 0.00 & 0.00 & 0.00 & 0.00 & 0.00\\
         \multicolumn{1}{r}{4-2} & 0.20 & 0.30 & 0.25 & 0.32 & 0.19 & 0.24 & 0.38 & 0.29 & 0.32 & 0.42 & 0.23 & 0.33\\
         5 Humor & 0.51 & 0.59 & 0.55 & 0.55 & 0.44 & 0.49 & 0.47 & 0.42 & 0.44 & 0.68 & 0.63 & 0.66\\
         6 Other & 0.00 & 0.00 & 0.00 & 0.00 & 0.00 & 0.00 & 0.00 & 0.00 & 0.00 & 0.00 & 0.00 & 0.00\\ 
         \multicolumn{1}{r}{6-1} & 0.45 & 0.56 & 0.51 & 0.38 & 0.35 & 0.36 & 0.65 & 0.54 & 0.59 & 0.70 & 0.76 & 0.73\\
         \multicolumn{1}{r}{6-2} & 0.42 & 0.39 & 0.41 & 0.41 & 0.41 & 0.41 & 0.45 & 0.27 & 0.33 & 0.50 & 0.56 & 0.54\\
         \midrule
         Macro Avg & 0.38 & 0.43 & 0.39 & 0.42 & 0.42 & 0.42 & 0.46 & 0.43 & 0.44 & 0.49 & 0.48 & 0.48\\
         Weighted Avg & 0.49 & 0.51 & 0.50 & 0.53 & 0.54 & 0.53 & 0.54 & 0.54 & 0.54 & 0.57 & 0.56 & 0.56\\
         \bottomrule
    \end{tabular}
    \label{sub_category}
\end{table*}

\section{Observations of the distribution of labels in chatbot usage dataset}\label{label_ditributions_des}
Figure~\ref{label_statistics} illustrates the label distribution across different languages within the Chatbot usage dataset. 
After annotating approximately 500 interaction data samples for each language, distinct trends became evident among diverse language communities.
For instance, subcategory 1-1 (Code \& Engineering) displayed a higher proportion within the \texttt{zh} language community. 
This suggests that a significant number of users within this community, possibly bypassing national regulations to use the platform, possess expertise in engineering and technical fields. 
Subcategory 1-2 (Politics \& Economy) was notably prevalent in the \texttt{tr} language community, while such inquiries were scarcely observed within the \texttt{ja} community. 
This distribution could mirror the contrasting levels of electoral engagement between Turkey and Japan, which rank 29th and 158th, respectively, in global voting rate rankings~\cite{vote_ranking}. 
The usage may reflect the respective societal interests in political and economic matters.
Analyzing the ratios between subcategories 3-1 and 3-2 provides insightful findings. 
Communities such as \texttt{tr}, \texttt{id}, and \texttt{nl} exhibited a higher ratio for 3-1 (Business Support), indicating a stronger inclination toward business-related interactions.
Conversely, communities such as \texttt{ja}, \texttt{pt}, \texttt{it}, and \texttt{ru} displayed a higher ratio for 3-2 (Creative Support), suggesting a tendency to use chatbots for creative endeavors. 
Category 4 serves as an indicator of whether users perceive chatbots as conversational partners. 
The data suggests a pronounced tendency within the \texttt{ja} and \texttt{pt} communities to engage with chatbots in a conversational manner.
Conversely, \texttt{es}, \texttt{de}, and \texttt{ko} communities demonstrated a lower count of tweets in this category, potentially reflecting a preference to use chatbots more for information retrieval than as conversational partners.
These observations provide a nuanced understanding of how diverse cultural and linguistic backgrounds may influence user interactions with chatbots, resulting in varied usage patterns across different thematic categories.

\section{Implementation details}

In our topic modeling analysis, as discussed in Sections 5 and 6, we utilized the BERTopic method based on version 0.15.0 of the Python framework~\cite{grootendorst2022bertopic_code}. 
The framework's default settings were used for most components. 
The document embeddings were generated using the pre-trained SentenceTransformer model, specifically the ``all-MiniLM-L6-v2'' variant~\cite{multilingual_sentence}. 
Since the model is pre-trained, no additional training epochs or fine-tuning of hyperparameters for the embedding process was required.
For dimensionality reduction, we utilized Uniform Manifold Approximation and Projection (UMAP)~\cite{mcinnes2018umap}. 
The UMAP settings included a neighborhood size of 15, a projection into 5 dimensions, and the use of a cosine similarity metric.
The clustering process was conducted using Hierarchical Density-Based Spatial Clustering of Applications with Noise (HDBSCAN), a robust density-based clustering algorithm~\cite{campello2013density}. 
We set the minimum cluster size to 10, the minimum core sample size to 3, and used the Euclidean distance metric as a similarity metric.
The combination of BERTopic's default settings, UMAP for dimensionality reduction and HDBSCAN for clustering, allowed us to generate interpretable and meaningful topic clusters from the text data, providing coherent insights into the underlying themes.

For the chatbot usage classification task, we implemented four distinct models as outlined in Section 6: TF-IDF + Logistic Regression, GPT-3, BERT, and DeBERTa.

\subsubsection{TF-IDF + Logistic Regression}
We exploit the TF-IDF feature extractor available in the sklearn library~\cite{pedregosa2011scikit}. 
This method computes term frequency-inverse document frequency (TF-IDF) values from the extracted text, which serve as input features for a logistic regression model. 
We opted for the LBFGS solver, also provided by sklearn, for the logistic regression implementation.

\subsubsection{GPT-3}
We leveraged the DaVinci variant of GPT-3~\cite{brown2020language}, accessed via OpenAI's API. 
For fine-tuning, the chatbot's usage text along with the correct labels served as the prompt. The response generated by the fine-tuned model was treated as the classification output, with the first string of the generated text being used as output.

\subsubsection{BERT}
We fine-tuned a pre-trained BERT model~\cite{bert_uncased} for 15 epochs. 
We set the batch size to 32 for the training set and 200 for the validation set.
We set the learning rate to $2 \times 10^{-5}$, employing the Adam optimizer~\cite{kingma2014adam} for weight adjustments.

\subsubsection{DeBERTa}
DeBERTa enhances the BERT architecture by differentiating between content and positional encodings through specialized attention mechanisms. 
This modification leads to notable improvements in task-specific performances. 
We fine-tuned a pretrained DeBERTa model~\cite{microsoft_debert} for 15 epochs.

\section{Classification of all categories}\label{all_classification}
We conducted an experiment involving the classification of all categories, encompassing both main categories and subcategories, resulting in a total of 19 categories. 
In this classification task, we treated main categories and subcategories as distinct classes for model training. 
For example, main Category 4 served as a class for all samples within main Category 4 that did not correspond to subcategories 4-1 and 4-2.
We employed four models (TF-IDF + Logistic Regression, GPT-3, BERT, and DeBERTa), as discussed in Section 6.3.3.

Table~\ref{sub_category} presents the accuracy of the classification.  
Across all models, the F1 score tended to improve with an increase in the number of samples, while it diminished with fewer samples (e.g., subcategory 4-1 or main category 6), demonstrating a proportional relationship with the sample count.
Logistic Regression, while achieving higher accuracy in classes with fewer samples compared to other models (e.g., subcategory 4-1), exhibited lower overall accuracy compared to deep learning-based models. 
Similar to the classification task involving only main categories (as described in Section 6.3.3), the DeBERTa model achieved the highest accuracy, although the margin of difference in accuracy between DeBERTa and other models was not substantial.
In this task, the F1 score for each model remained around 0.5. 
For practical applications, measures such as increasing the number of samples may be necessary to enhance performance.

\section{Classification of chatbot usage in WildChat dataset}

In RQ3, we explored chatbot usage across different linguistic communities based on screenshots posted on Twitter. 
To complement this analysis, we applied the same chatbot usage classification to the publicly available WildChat dataset, consisting of over 1 million user ChatGPT interaction logs~\cite{zhao2024wildchat}. 
This dataset includes interactions between person and ChatGPT in 68 languages with user locations, IP addresses, and time stamps, which enables a comprehensive analysis of user behavior and chatbot interactions.
The results are expected to align with RQ3's purpose of understanding chatbot usage across different language communities.

Figure~\ref{wildchat_result} shows the results of chatbot usage classification applied to the WildChat dataset, broken down by language.
In language communities such as \texttt{nl}, \texttt{pt}, and \texttt{ru}, ``Search'' (category 1) usage is predominant. 
Meanwhile, in \texttt{de}, \texttt{en}, and \texttt{ja}, the ``Support'' category (category 3) represents the majority of interactions. 
Categories like ``Dialogue'' (category 4) and ``Humor'' (category 5) are relatively uncommon across all language groups, although ``Dialogue'' accounts for approximately 7\% in \texttt{en}, which is notably higher compared to other languages. 


However, these results deviate from the label distributions observed in the RQ3 analysis, as shown in Appendix M.
This discrepancy arises from differences in the dataset characteristics.
In RQ3, we focused on user-generated posts shared on Twitter, where users tend to display use cases they wish to share publicly rather than regular usage patterns, resulting in diverges from real-world use.
The WildChat dataset also faces challenges in accurately capturing real-world usage across languages.
The dataset was collected by providing free access to ChatGPT through APIs, with users opting into anonymous data collection. 
As a result, there is inherent bias in user demographics, as the dataset predominantly reflects interactions from users who are comfortable using API-based interfaces, excluding a broader demographic of users. 
Moreover, upon examining the dataset, we found substantial noise: for instance, some interactions labeled as Japanese contain Chinese text, and over half of the 15,000 interactions labeled as German start with the prompt ``Gib mir nur 10 Keywords...''.
These issues highlight significant biases within the dataset.
Despite the limitations of the WildChat dataset, we expect its chatbot usage distribution to be a complementary resource to RQ3. 
The different perspectives provided by this dataset, alongside findings in RQ3, offer a more comprehensive understanding of chatbot usage across linguistic communities.

\begin{table*}
    \centering
    \small
    \caption{Label counts and percentages by language in WildChat dataset applied to chatbot usage classification model based on DeBERTa.}
    \begin{tabular}{crrrrrrr}
        \toprule
        \multirow{2}{*}{Lang} & \multicolumn{7}{c}{\textbf{Label (Percentage (Count)})} \\ 
        & 1 Search & 2 Preference  & 3 Support & 4 Dialogue & 5 Humor & 6 Other & No. of samples \\ \midrule
        \texttt{ar} & 33.41\% (2,660) & 23.05\% (1,835) & 36.39\% (2,897) & 4.03\% (321) & 0.34\% (27) & 2.78\% (221) & 7,961\\ 
        \texttt{de} & 26.77\% (4,046) & 5.26\% (795) & 65.86\% (9,955) & 1.08\% (163) & 0.34\% (52) & 0.69\% (105) & 15,166\\ 
        \texttt{en} & 29.13\% (80,735) & 10.70\% (29,664) & 49.80\% (138,032) & 7.88\% (21,853) & 0.88\% (2,433) & 1.61\% (4,458) &  277,175\\ 
        \texttt{es} & 37.41\% (4,487) & 16.15\% (1,937) & 39.37\% (4,723) & 4.39\% (526) & 0.50\% (60) & 2.18\% (262) &  11,995\\ 
        \texttt{fr} & 44.73\% (7,251) & 20.68\% (3,352) & 29.98\% (4,860) & 2.76\% (448) & 0.46\% (75) & 1.38\% (223) & 16,209\\ 
        \texttt{id} & 53.52\% (532) & 15.79\% (157) & 22.54\% (224) & 1.21\% (12) & 0.10\% (1) & 6.84\% (68) & 994\\ 
        \texttt{it} & 42.88\% (1,400) & 19.17\% (626) & 29.98\% (979) & 4.93\% (161) & 0.31\% (10) & 2.73\% (89) & 3,265\\ 
        \texttt{ja} & 20.21\% (540) & 9.21\% (246) & 59.06\% (1,578) & 9.28\% (248) & 0.71\% (19) & 1.53\% (41) & 2,672\\ 
        \texttt{ko} & 45.58\% (201) & 20.41\% (90) & 23.13\% (102) & 7.26\% (32) & 0.45\% (2) & 3.17\% (14) & 441\\ 
        \texttt{nl} & 55.54\% (817) & 15.91\% (234) & 22.98\% (338) & 2.79\% (41) & 0.61\% (9) & 2.18\% (32) & 1,471\\ 
        \texttt{pt} & 49.57\% (3,299) & 15.01\% (999) & 26.16\% (1,741) & 4.09\% (272) & 3.01\% (200) & 2.16\% (144) & 6,655\\ 
        \texttt{ru} & 48.85\% (23,932) & 14.28\% (6,998) & 30.57\% (14,975) & 3.80\% (1,864) & 0.66\% (321) & 1.85\% (904) & 49,994\\ 
        \texttt{tr} & 45.12\% (1,342) & 15.84\% (471) & 32.18\% (957) & 4.40\% (131) & 0.30\% (9) & 2.15\% (64) & 2,974\\
        \texttt{zh} & 42.83\% (42,199) & 16.39\% (16,148) & 36.27\% (35,743) & 2.13\% (2,103) & 0.39\% (388) & 1.99\% (1,957) & 98,538\\ \bottomrule
    \end{tabular}
    \label{wildchat_result}
\end{table*}

\section{Generative AI Percepiton in VKontakte (VK)}

We focused on understanding perceptions of generative AI tools across various linguistic communities by analyzing Twitter, a platform widely used by users from different language communities. 
By focusing on a single platform, we could highlight both language-specific trends and cross-linguistic patterns.
However, Twitter’s primary user base consists mostly of English and Japanese speakers, while users of other languages may primarily engage with different platforms.
To address this gap, we extended our research to VKontakte (VK), a social media platform predominantly used by Russian communities. 
By conducting a similar analysis on VK, we aim to reveal how perceptions of generative AI differ between VK and Twitter, shedding light on platform-specific differences in user opinions.

\noindent \textbf{Data}: 
We collected data from VK using the platform’s API, applying the same set of keywords used in our analysis.
This approach ensured consistency in identifying relevant posts about generative AI tools across platforms. 
As a result, we gathered a total of 40,866 posts, as summarized in Table~\ref{vk_table}.

\begin{table}
    \centering
    \caption{Number of posts in VK across generative AI tool.}
    \label{vk_table}
    \begin{tabular}{llr}
        \toprule
        Category & Generative AI tool & No. of posts \\ \midrule
        \multirow{3}{*}{Chat}  & ChatGPT & 15,766 \\ 
        & Bing Chat & 66 \\ 
        & Perplexity AI & 10 \\ \midrule
        \multirow{6}{*}{Image} & DALL $\cdot$ E & 5,315 \\
        & Stable Diffusion & 3,408\\
        & MidJourney & 14,977\\
        & Craiyon & 99 \\
        & DreamStudio & 10 \\ \midrule
        Code & Github Copilot & 18\\ \midrule
        \multirow{2}{*}{Model} & GPT-3 & 591 \\
        & GPT-4 & 606\\ 
        \bottomrule
    \end{tabular}
\end{table}

\begin{figure}[t]
    \centering
    \includegraphics[width=8cm]{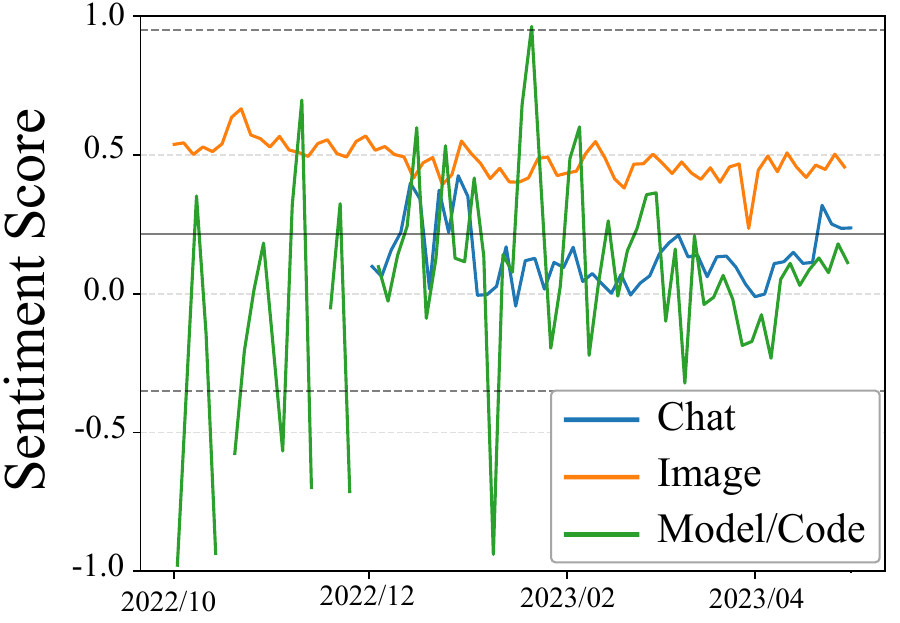}
    \caption{
    Time series of sentiment score of VK posts in \texttt{ru}. 
    The x-axis represents time, and the y-axis represents the sentiment score.
    The sentiment score for generative AI tools associated with Chat is represented by the blue time series, while the orange corresponds to Image, and green to Model/Code.
    The solid gray line represents the average sentiment score, while the gray dotted lines indicate the 75th and 25th percentiles of sentiment scores from typical posts.
    }
    \label{ru_vk_sentiment}
\end{figure}

\noindent \textbf{Sentiment (RQ1)}: 
We applied the Russian sentiment classification model used in RQ1 to VK posts to examine the perception of generative AI tools.
The sentiment analysis revealed an overall average sentiment score of 0.220, with the following score for specific tools: 0.123 for Chat tools, 0.472 for Image tools, and 0.010 for Model tools, as shown in Figure~\ref{ru_vk_sentiment}.
These findings are consistent with those from Twitter data (Section 4), where Image tools also received the highest sentiment score, and Model tools received the lowest.
However, one key difference between the VK and Twitter data is that, while on Twitter all sentiment scores were below the overall average, the sentiment score for Image tools on VK was notably higher than the average.
This suggests that \texttt{ru} users on VK are more engaged with and show greater interest in image generation tools compared to Twitter users.
Indeed, upon closer inspection of the actual posts, there is a high level of activity surrounding discussions on image generation, further supporting the idea that VK users have a stronger focus on and enthusiasm for Image tools than their Twitter counterparts.

\begin{figure}[t]
    \centering
    \caption{Top 10 words in posts in VK more likely to be used during three periods by odds ratio, except generative AI tool’ names.}
    \includegraphics[width=9cm]{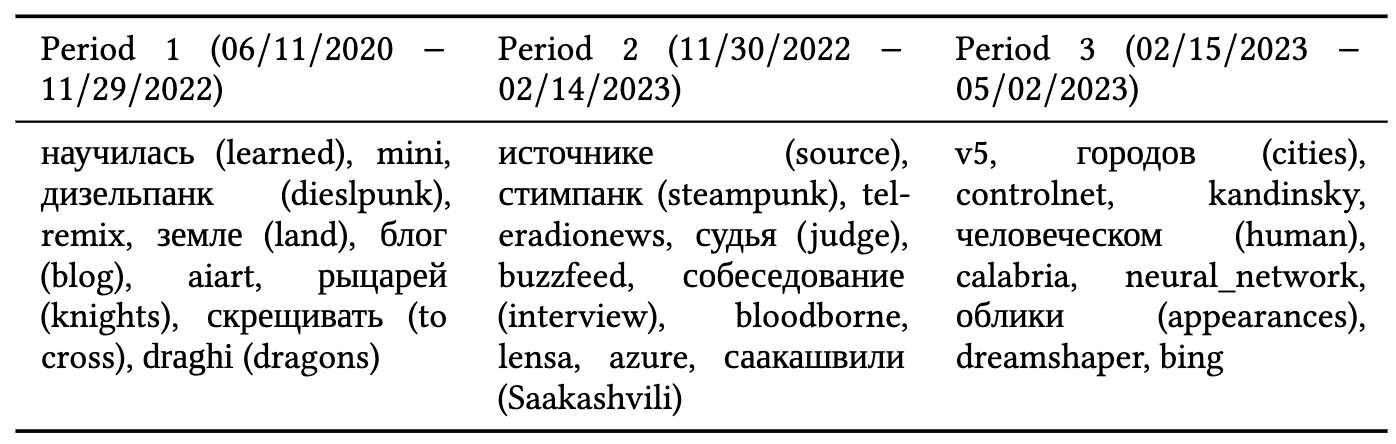}
    \label{odds_vk}
\end{figure}

\noindent \textbf{Odds ratio analysis  (RQ2)}:
As part of our investigation into RQ2, we applied the same analysis to VK posts to examine the content related to generative AI tools.
Similar to our previous approach in RQ2, we utilized an odds ratio analysis to quantitatively measure the frequency of specific terms used in these discussions.
This allows us to identify the most common words and gain insights into the dominant topics within VK posts about generative AI.

The results of the odds ratio analysis are presented in Figure~\ref{odds_vk}.
Similar to the trends observed on Twitter, the word ``buzzfeed'' was particularly prominent during Period 2.
Additionally, in Period 3, there was a significant focus on the technical aspects of generative AI,  with words such as``neural\_network'', and ``v5'' appearing.
However, there are also unique trends specific to VK. 
For example, in Period 1, words like ``dieselpunk,'' ``knights'' and ``dragons'' frequently appeared, suggesting these words were commonly used to generate images with AI tools.
This suggests that VK users were particularly engaged with the creative applications of generative AI during this period.
Furthermore, in Period 3, the word ``human'' became more prominent, indicating a shift in discussions toward evaluating the human-like qualities of generative AI.
This reflects an increased interest among VK users in how AI tools emulate human characteristics.
These findings highlight both the similarities and differences in how generative AI tools are discussed on VK compared to other platforms like Twitter.
In particular, VK users appear to demonstrate a distinct focus on creative and humanistic aspects of AI.

\begin{figure}[t]
    \centering
    \caption{Keywords for the top ten topics and topic number in terms of number of posts by BERTopic.}
    \includegraphics[width=8.5cm]{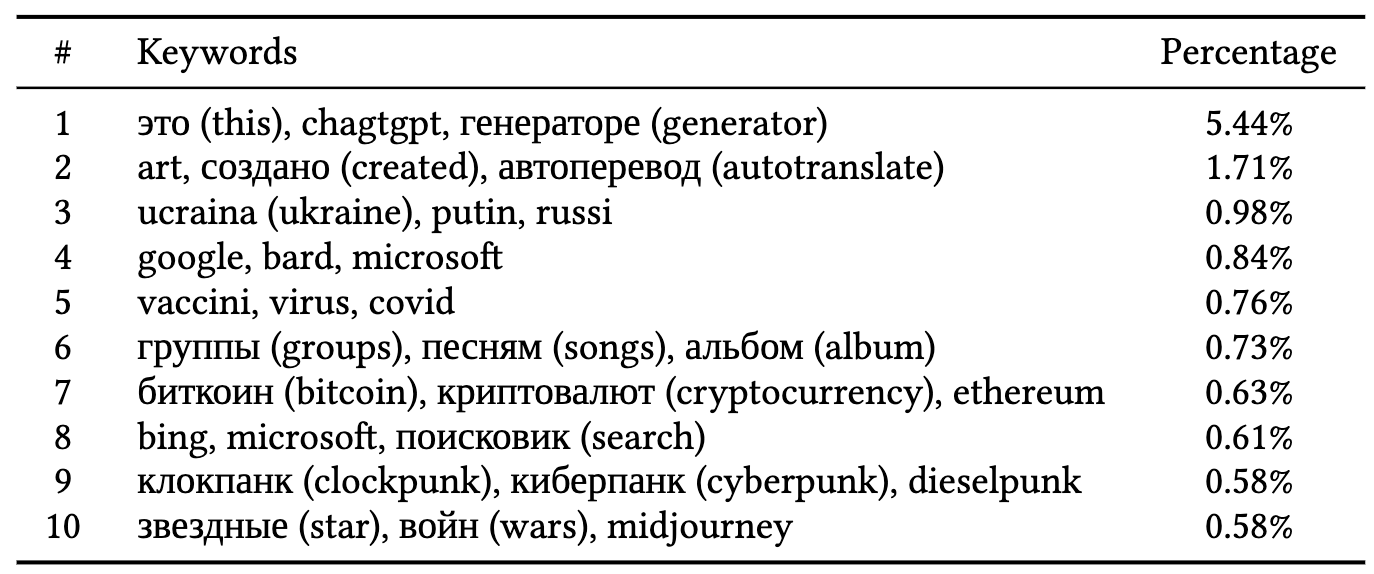}
    \label{topic_model_vk}
\end{figure}

\noindent \textbf{Topic Model  (RQ2)}:
As part of our investigation into RQ2, we applied the BERTopic model to analyze the content of VK posts related to generative AI tools.
The results revealed that 39.4\% of the data were considered outliers, and the model identified 498 distinct topics.
Figure~\ref{topic_model_vk} presents the top 10 topics.
Similar to our findings on Twitter, topics related to current global events, such as the Ukraine conflict (Topic 3) and discussions surrounding cryptocurrency (Topic 7), ranked highly in VK as well.
However, unlike Twitter, where discussions on the Italy ban frequently appeared among the top topics across multiple language communities, this topic did not rank as highly on VK.
One notable difference specific to VK is the prominence of creative related topics, such as Topic 9 and Topic 10, which are centered on discussions of art, design, and creative works generated using AI.
This suggests that VK users engage more actively with generative AI for creative purposes, highlighting a distinct difference in how the platform's user base interacts with these tools compared to Twitter.

\noindent \textbf{Chatbot usage (RQ3)}:
We finally classified ChatGPT screenshots from VK posts containing images to examine chatbot usage, following a similar approach as in RQ3.
While ChatGPT was mentioned in approximately 11.7\% of posts containing screenshots on Twitter's \texttt{ru} users (2,181 out of 18,531), on VK, it appeared in only 8.9\% of posts (1406 out of 15,788.)
Although there were fewer screenshots related to ChatGPT on VK, there was a noticeable increase in posts containing images generated by tools like MidJourney. 
This suggests that VK users are more actively sharing AI-generated images compared to their Twitter counterparts.

When classifying chatbot usage, we found no significant differences from Twitter.
On Twitter, the distribution of chatbot usage was as follows:
1. Search: 49.6\% (1,082),
2. Preference: 14.4\% (315),
3. Support: 22.2\% (484),
4. Dialogue: 4.3\% (134),
5. Humor: 6.1\% (73),
6. Other: 4.3\% (93).

In comparison, on VK, the distribution was:
1. Search: 44.8\% (630),
2. Preference: 16.1\% (226),
3. Support: 21.0\% (295),
4. Dialogue: 5.9\% (83),
5. Humor: 2.7\% (40),
6. Other: 9.3\% (132).

These results indicate that while VK users share fewer ChatGPT-related screenshots compared to Twitter users, the overall pattern of chatbot usage is similar across both platforms. 
The most common usage category remains ``1. Search,'' followed by ``3. Support'' and ``2. Preference,'' with slight variations in less frequent categories like ``4. Dialogue'' and ``5. Humor.''

\begin{table*}
  \scriptsize
  \caption{Tweet volume and IntI by language and generative AI tools.}
  \label{lang_service}
  \scalebox{0.92}{
  \begin{tabular}{clrr|clrr|clrr}
    \toprule
    Language & Service & Number of tweets & IntI & Language & Service & Number of tweets & IntI & Language & Service & Number of tweets & IntI\\ \midrule
    \multirow{12}{*}{en} & Chat GPT& 2,430,739 (59.97\%)& 1.00 & 
    \multirow{12}{*}{ja} & Chat GPT& 969,188 (66.20\%)& 0.51& 
    \multirow{12}{*}{es} & Chat GPT& 302,702 (77.59\%)& 0.53 
    \\
    & Bing Chat & 18,052 (0.25\%)& 1.00& 
    & Bing Chat & 3,619 (0.24\%)& 0.25& 
    & Bing Chat & 1604 (0.41\%)& 0.38
    \\
    & Perplexity AI & 2,211 (0.05\%)& 1.00 & 
    & Perplexity AI & 1,818 (0.12\%)& 1.05 & 
    & Perplexity AI & 314 (0.01\%)& 0.61 
    \\
    & DALL·E & 331,877 (8.19\%) & 1.00 & 
    & DALL·E & 13,363 (0.91\%) & 0.05 & 
    & DALL·E & 23,909 (6.12\%) & 0.31 
    \\
    & Stable Diffusion & 317,451 (7.8\%)& 1.00 & 
    & Stable Diffusion & 205,358 (14.02\%)& 0.82& 
    & Stable Diffusion & 12,641 (3.24\%)& 0.17 
    \\
    & Midjourney & 582,619 (14.37\%)& 1.00 & 
    & Midjourney & 182,219 (12.44\%)& 0.40 & 
    & Midjourney & 25,792 (6.61\%)& 0.19 
    \\
    & Craiyon & 64,638 (1.59\%)& 1.00 & 
    & Craiyon & 6,519 (0.44\%)& 0.13 & 
    & Craiyon & 3,861 (0.99\%)& 0.25 
    \\
    & DreamStudio & 9,953 (0.24\%) & 1.00 & 
    & DreamStudio & 3,961 (0.27\%) & 0.51 & 
    & DreamStudio & 548 (0.15\%) & 0.23 
    \\
    & Github Copilot & 29,735 (0.72\%) & 1.00 & 
    & Github Copilot & 12,312 (0.84\%) & 0.53 & 
    & Github Copilot & 2,698 (0.69\%) & 0.39 
    \\
    & GPT-3 & 284,750 (7.04\%) & 1.00 & 
    & GPT-3 & 33,310 (2.27\%) & 0.15 &  
    & GPT-3 & 19,888 (5.10\%) & 0.30 
    \\
    & GPT-3.5 & 19,491(0.47\%) & 1.00 & 
    & GPT-3.5 & 1,301 (0.08\%) & 0.08 & 
    & GPT-3.5 & 1,272 (0.32\%) & 0.28 
    \\
    & GPT-4 & 279,542 (6.89\%) & 1.00 &  
    & GPT-4 & 85,642 (5.85\%) & 0.39 & 
    & GPT-4 & 20,536 (5.26\%) & 0.31 
    \\ \midrule

    \multirow{12}{*}{fr} & Chat GPT& 164,716 (65.92\%)& 1.18& 
    \multirow{12}{*}{pt} & Chat GPT& 165,874 (85.04\%)& 0.33& 
    \multirow{12}{*}{zh} & Chat GPT& 109,754 (81.47\%)& 2.26
    \\
    & Bing Chat & 574 (0.23\%)& 0.55& 
    & Bing Chat & 747 (0.38\%)& 0.20& 
    & Bing Chat & 527 (0.39\%)& 1.46 
    \\
    & Perplexity AI & 68 (0.03\%)& 0.54 & 
    & Perplexity AI & 58 (0.03\%)& 0.13 & 
    & Perplexity AI & 53 (0.04\%)& 0.15 
    \\
    & DALL·E & 10,547 (4.22\%) & 0.55 & 
    & DALL·E & 6,794 (3.48\%) & 0.10 & 
    & DALL·E & 915 (0.68\%) & 0.14 
    \\
    & Stable Diffusion & 12,507 (5.01\%)& 0.69 & 
    & Stable Diffusion & 2,893 (1.48\%)& 0.04 & 
    & Stable Diffusion & 4,526 (3.35\%)& 0.72
    \\
    & Midjourney & 54,021 (21.62\%)& 1.61 & 
    & Midjourney & 11,580 (5.93\%)& 0.10 & 
    & Midjourney & 8,197 (6.08\%)& 0.71 
    \\ 
    & Craiyon & 1,313 (0.52\%)& 0.35 & 
    & Craiyon & 1,095 (0.56\%)& 0.08 & 
    & Craiyon & 41 (0.03\%)& 0.03
    \\
    & DreamStudio & 224 (0.09\%) & 0.39 & 
    & DreamStudio & 142 (0.07\%) & 0.07 & 
    & DreamStudio & 36 (0.02\%) & 0.18 
    \\
    & Github Copilot & 1,002 (0.40\%) & 0.59 & 
    & Github Copilot & 1,273 (0.65\%) & 0.21 & 
    & Github Copilot & 1,104 (0.81\%) & 1.86 
    \\
    & GPT-3 & 8,297 (3.32\%) & 0.51 & 
    & GPT-3 & 3,900 (1.99\%) & 0.07 &  
    & GPT-3 & 3,422 (2.54\%) & 0.60 
    \\
    & GPT-3.5 & 922 (0.37\%) & 0.83 & 
    & GPT-3.5 & 363 (0.19\%) & 0.09 & 
    & GPT-3.5 & 512 (0.38\%) & 1.32 
    \\
    & GPT-4 & 12,056 (4.83\%) & 0.75 & 
    & GPT-4 & 6,688 (3.43\%) & 0.12 & 
    & GPT-4 & 16,882 (12.53\%) & 3.03
    \\ \midrule

    \multirow{12}{*}{de} & Chat GPT& 72,413 (76.95\%)& 1.22 & 
    \multirow{12}{*}{tr} & Chat GPT& 37,916 (65.81\%)& 0.19 & 
    \multirow{12}{*}{id} & Chat GPT& 35,267 (69.04\%)& 0.13 
    \\
    & Bing Chat & 277 (0.29\%)& 0.63& 
    & Bing Chat & 157 (0.28\%)& 0.11 & 
    & Bing Chat & 197 (0.39\%)& 0.10
    \\
    & Perplexity AI & 34 (0.04\%)& 0.08 & 
    & Perplexity AI & 9 (0.01\%)& 0.01 & 
    & Perplexity AI & 609 (1.19\%)& 0.30 
    \\
    & DALL·E & 5,978 (6.35\%) & 0.74 & 
    & DALL·E & 2,707 (4.70\%) & 0.10 & 
    & DALL·E & 2,069 (4.05\%) & 0.06 
    \\
    & Stable Diffusion & 3,318 (3.53\%)& 0.43 & 
    & Stable Diffusion & 1,032 (1.79\%)& 0.04 & 
    & Stable Diffusion & 3,129 (6.13\%)& 0.09 
    \\
    & Midjourney & 6,419 (6.82\%)& 0.45 & 
    & Midjourney & 6,838 (11.87\%)& 0.14 & 
    & Midjourney & 5,547 (10.86\%)& 0.08
    \\
    & Craiyon & 1,846 (1.96\%)& 1.17& 
    & Craiyon & 181 (0.31\%)& 0.03& 
    & Craiyon & 644 (1.26\%)& 0.08
    \\
    & DreamStudio & 106 (0.11\%) & 0.44 & 
    & DreamStudio & 14 (0.02\%) & 0.02 & 
    & DreamStudio & 71 (1.39\%) & 0.06 
    \\
    & Github Copilot & 208 (0.22\%) & 0.29 & 
    & Github Copilot & 611 (1.06\%) & 0.25 & 
    & Github Copilot & 525 (1.03\%) & 0.16 
    \\
    & GPT-3 & 5,093 (5.41\%) & 0.74 & 
    & GPT-3 & 3,048 (5.29\%) & 0.13 &  
    & GPT-3 & 2,555 (5.01\%) & 0.02 
    \\
    & GPT-3.5 & 182 (0.20\%) & 0.38 & 
    & GPT-3.5 & 285 (0.32\%) & 0.18 & 
    & GPT-3.5 & 222 (0.42\%) & 0.10 
    \\
    & GPT-4 & 4,572 (4.86\%) & 0.68 & 
    & GPT-4 & 8,712 (15.12\%) & 0.38 & 
    & GPT-4 & 3,085 (6.04\%) & 0.10
    \\ \midrule

    \multirow{12}{*}{ar} & Chat GPT& 41,267 (86.92\%)& 0.70 & 
    \multirow{12}{*}{it} & Chat GPT& 24,944 (59.92\%)& 0.41 & 
    \multirow{12}{*}{ru} & Chat GPT& 18,531 (63.89\%)& 0.50 
    \\
    & Bing Chat & 199 (0.42\%)& 0.27& 
    & Bing Chat & 113 (0.27\%)& 0.25 & 
    & Bing Chat & 40 (0.14\%)& 0.15 
    \\ 
    & Perplexity AI & 14 (0.03\%)& 0.03 & 
    & Perplexity AI & 24 (0.06\%)& 0.05 & 
    & Perplexity AI & 7 (0.03\%)& 0.03 
    \\
    & DALL·E & 634 (1.33\%) & 0.08 & 
    & DALL·E & 8,682 (20.86\%) & 1.05 & 
    & DALL·E & 1,198 (4.13\%) & 0.24 
    \\
    & Stable Diffusion & 187 (0.39\%) & 0.03 & 
    & Stable Diffusion & 1,389 (3.33\%)& 0.17 & 
    & Stable Diffusion & 993 (3.42\%)& 0.21 
    \\
    & Midjourney & 3,039 (6.40\%)& 0.22 & 
    & Midjourney & 3,522 (8.46\%)& 0.24 & 
    & Midjourney & 5,883 (20.28\%)& 0.67
    \\
    & Craiyon & 36 (0.08\%)& 0.02 & 
    & Craiyon & 4,085 (9.81\%)& 2.51 & 
    & Craiyon & 122 (0.42\%)& 0.12 
    \\
    & DreamStudio & 10 (0.02\%) & 0.04 & 
    & DreamStudio & 51 (0.12\%) & 0.20 & 
    & DreamStudio & 15 (0.05\%) & 0.10 
    \\
    & Github Copilot & 208 (0.44\%) & 0.29 & 
    & Github Copilot & 33 (0.07\%) & 0.04 & 
    & Github Copilot & 202 (0.70\%) & 0.45 
    \\
    & GPT-3 & 1,536 (2.24\%) & 0.22 & 
    & GPT-3 & 1,199 (2.88\%) & 0.17 & 
    & GPT-3 & 1,554 (5.36\%) & 0.36 
    \\
    & GPT-3.5 & 124 (0.25\%) & 0.26 & 
    & GPT-3.5 & 218 (0.56\%) & 0.44 & 
    & GPT-3.5 & 129 (0.44\%) & 0.44 
    \\
    & GPT-4 & 3,564 (7.50\%) & 0.53 & 
    & GPT-4 & 3,416 (8.21\%) & 0.49 & 
    & GPT-4 & 1,934 (6.67\%) & 0.46 
    \\ \midrule

    \multirow{12}{*}{ko} & Chat GPT& 19,845 (69.65\%)& 0.08& 
    \multirow{12}{*}{nl} & Chat GPT& 18,868 (67.26\%)& 0.63 & 
     & & &
    \\
    & Bing Chat & 194 (0.68\%)& 0.11 & 
    & Bing Chat & 76 (0.27\%)& 0.34 & 
    & & &
    \\
    & Perplexity AI & 14 (0.05\%)& 0.01 & 
    & Perplexity AI & 16 (0.05\%)& 0.07 & 
    & & &
    \\
    & DALL·E & 890 (3.12\%) & 0.03 & 
    & DALL·E & 1,299 (4.63\%) & 0.32 & 
    & & &
    \\
    & Stable Diffusion & 2,436 (8.55\%)& 0.08 & 
    & Stable Diffusion & 629 (2.24\%) & 0.16 & 
    &  & &
    \\
    & Midjourney & 968 (3.40\%) & 0.02& 
    & Midjourney & 2,140 (7.63\%)& 0.30 & 
    & & &
    \\
    & Craiyon & 146 (0.51\%)& 0.02 & 
    & Craiyon & 270 (0.96\%)& 0.34 & 
    & & &
    \\
    & DreamStudio & 6 (0.02\%) & 0.01 & 
    & DreamStudio & 65 (0.23\%) & 0.53 & 
    & & &
    \\
    & Github Copilot & 97 (0.97\%) & 0.04 & 
    & Github Copilot & 55 (0.20\%) & 0.15 & 
    & & &
    \\
    & GPT-3 & 1,625 (5.71\%) & 0.06& 
    & GPT-3 & 870 (3.10\%) & 0.25 & 
    & & &
    \\
    & GPT-3.5 & 158 (0.55\%) & 0.08 & 
    & GPT-3.5 & 25 (0.09\%) & 0.10 & 
    & & &
    \\
    & GPT-4 & 1,253 (12.54\%) & 0.05 & 
    & GPT-4 & 4,911 (17.51\%) & 1.42 & 
    & & &
    \\ 
  \bottomrule
\end{tabular}
}
\end{table*}

\begin{figure*}
\centering
  \caption{Top 10 words in tweets more likely to be used during three periods by odds ratio, except generative AI tool' names. 
  Red indicates words that appear in multiple languages and blue indicates words unique to a specific language.}
  \label{odds_ratio_table_appendix}
  \includegraphics[width=18cm]{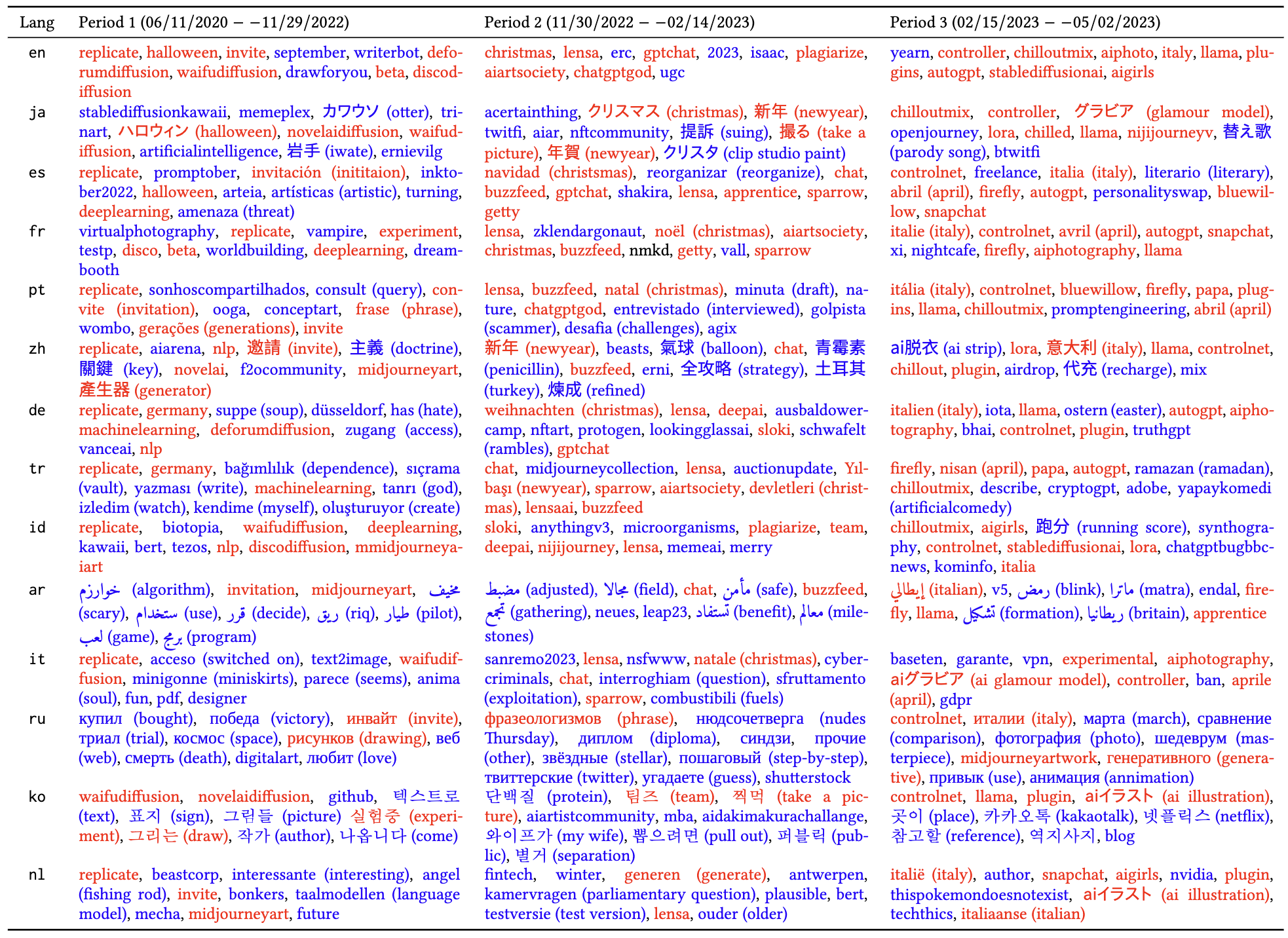}
\end{figure*}

\begin{figure*}
    \centering
    \includegraphics[width=18cm]{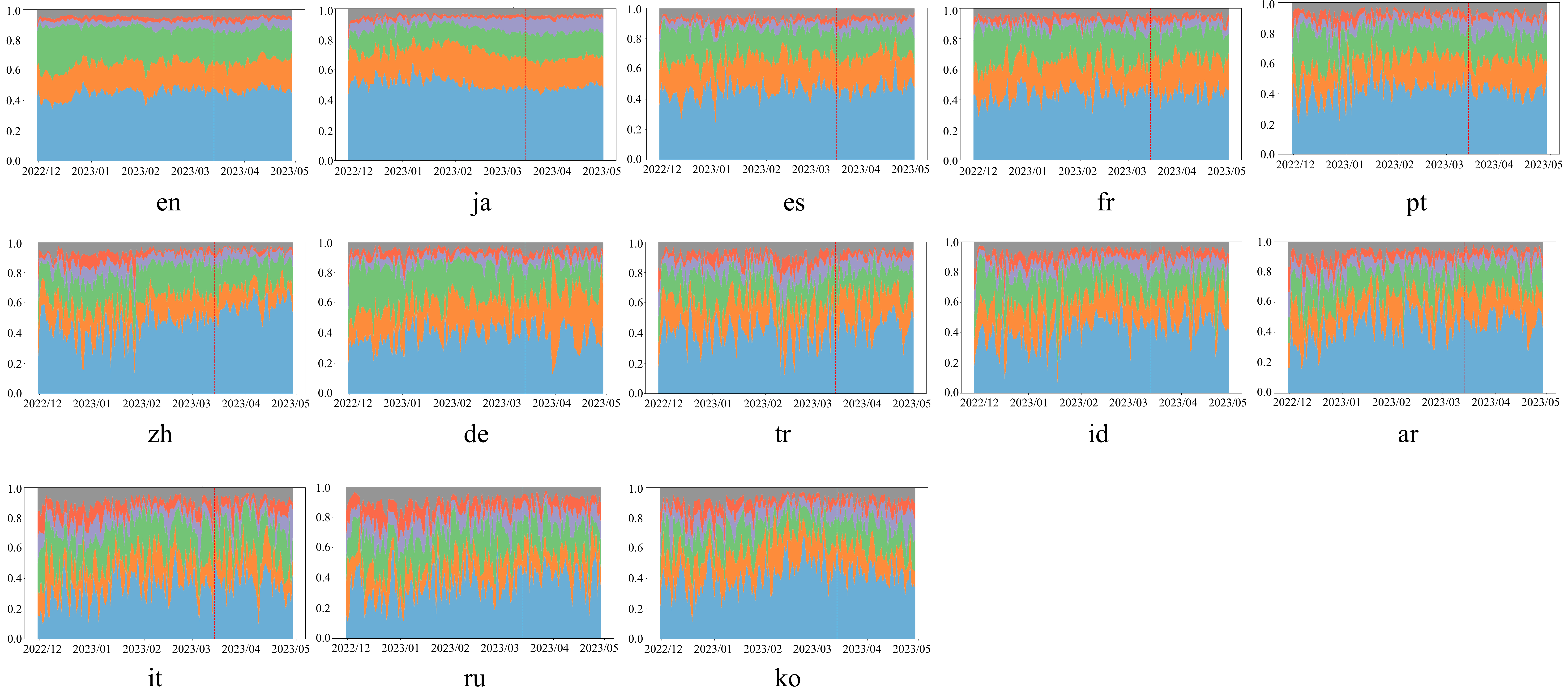}
    \caption{The temporal evolution of chatbot usage across various categories within 14 language communities. 
    The vertical red line serves as a reference point for the release date of GPT-4. 
    Each color corresponds to a different main category: blue for main Category 1, orange for main Category 2, green for main Category 3, purple for main Category 4, red for main Category 5, and gray for main Category 6.
    On the chart, the x-axis represents time, while the y-axis represents either the number or ratio of interactions in each category.
    Notably, the temporal evolution of chatbot usage in the \texttt{nl} community is so sparse that it is not included in this representation.}
    \label{timeseries_classification_appendix}
\end{figure*}

\end{document}